\documentclass{article}

\usepackage[nonatbib,preprint]{neurips_data_2023}

\usepackage[utf8]{inputenc} 
\usepackage[T1]{fontenc}    
\usepackage{hyperref}       
\usepackage{url}            
\usepackage{booktabs}       
\usepackage{amsfonts}       
\usepackage{nicefrac}       
\usepackage{xr}
\usepackage{microtype}      
\usepackage[dvipsnames,svgnames]{xcolor}
\usepackage{soul}         
\usepackage{amsmath,bm}
\usepackage{multirow}
\usepackage{array}
\usepackage{makecell}
\usepackage{graphicx}
\usepackage[small]{caption}
\usepackage{diagbox}
\usepackage{tabularx}
\usepackage{adjustbox}
\usepackage{bbding}
\usepackage{transparent}
\usepackage{subcaption}
\usepackage{amsfonts}

\usepackage{ulem}

\usepackage[noabbrev,nameinlink]{cleveref}
\crefname{property}{property}{Property}
\creflabelformat{property}{(#1)#2#3}
\creflabelformat{equation}{#1#2#3}

\newcolumntype{x}[1]{>{\centering\arraybackslash\hspace{0pt}}p{#1}}

\makeatletter
\newcommand{\thickhline}{%
    \noalign {\ifnum 0=`}\fi \hrule height 1.3pt
    \futurelet \reserved@a \@xhline
}

\title{$\mathbf{\mathbb{E}^{FWI}}$: Multiparameter Benchmark Datasets for \\ Elastic Full Waveform Inversion of \\ Geophysical Properties}

\author{%
  Shihang Feng$^{1,}$\thanks{Equal contribution}\\
  \And 
  Hanchen Wang$^{1, *}$\\
  \And 
  Chengyuan Deng$^{1, 2}$ \\
  \And
  Yinan Feng$^{1}$\\
  \And
  Yanhua Liu$^{1,3}$\\
  \And
  Min Zhu$^{1,4}$\\
  \And 
  Peng Jin$^{1,5}$\\
  \And 
  Yinpeng Chen$^{6}$\\
  \And 
  Youzuo Lin$^{1}$\\
   \And
\textnormal{$^1$Los Alamos National Laboratory \ ~~~
$^2$Rutgers University \
$^3$Colorado School of Mines \
} \\
$^4$University of Pennsylvania \
$^5$The Pennsylvania State University \
$^6$Microsoft \\
\texttt{\{shihang, hanchen.wang, charles.deng, ynf, yanhualiu, minzhu, pjin, ylin\}@lanl.gov}\\
\texttt{yiche@microsoft.com}
}

\begin{document}

\maketitle

\begin{abstract}

Elastic geophysical properties (such as P- and S-wave velocities) are of great importance to various subsurface applications like CO$_2$ sequestration and energy exploration~(e.g., hydrogen and geothermal). Elastic full waveform inversion (FWI) is widely applied for characterizing reservoir properties. In this paper, we introduce $\mathbf{\mathbb{E}^{FWI}}$, a comprehensive benchmark dataset that is specifically designed for elastic FWI. $\mathbf{\mathbb{E}^{FWI}}$ encompasses 8 distinct datasets that cover diverse subsurface geologic structures (flat, curve, faults, etc). The benchmark results produced by three different deep learning methods are provided. In contrast to our previously presented dataset (pressure recordings) for acoustic FWI~(referred to as \textsc{OpenFWI}), the seismic dataset in $\mathbf{\mathbb{E}^{FWI}}$ has both vertical and horizontal components. Moreover, the velocity maps in $\mathbf{\mathbb{E}^{FWI}}$ incorporate both P- and S-wave velocities. While the multicomponent data and the added S-wave velocity make the data more realistic, more challenges are introduced regarding the convergence and computational cost of the inversion. We conduct comprehensive numerical experiments to explore the relationship between P-wave and S-wave velocities in seismic data. The relation between  P- and S-wave velocities provides crucial insights into the subsurface properties such as lithology, porosity, fluid content, etc.  We anticipate that $\mathbf{\mathbb{E}^{FWI}}$ will facilitate future research on multiparameter inversions and stimulate endeavors in several critical research topics of carbon-zero and new energy exploration. All datasets, codes\footnote[1]{Codes will be released upon approval by Los Alamos National Laboratory and U.S. Department of Energy.} and relevant information can be accessed through our website at~\url{https://efwi-lanl.github.io/}.


\end{abstract}

\begin{figure*}[h!]
\centering
\includegraphics[width=1.0\columnwidth]{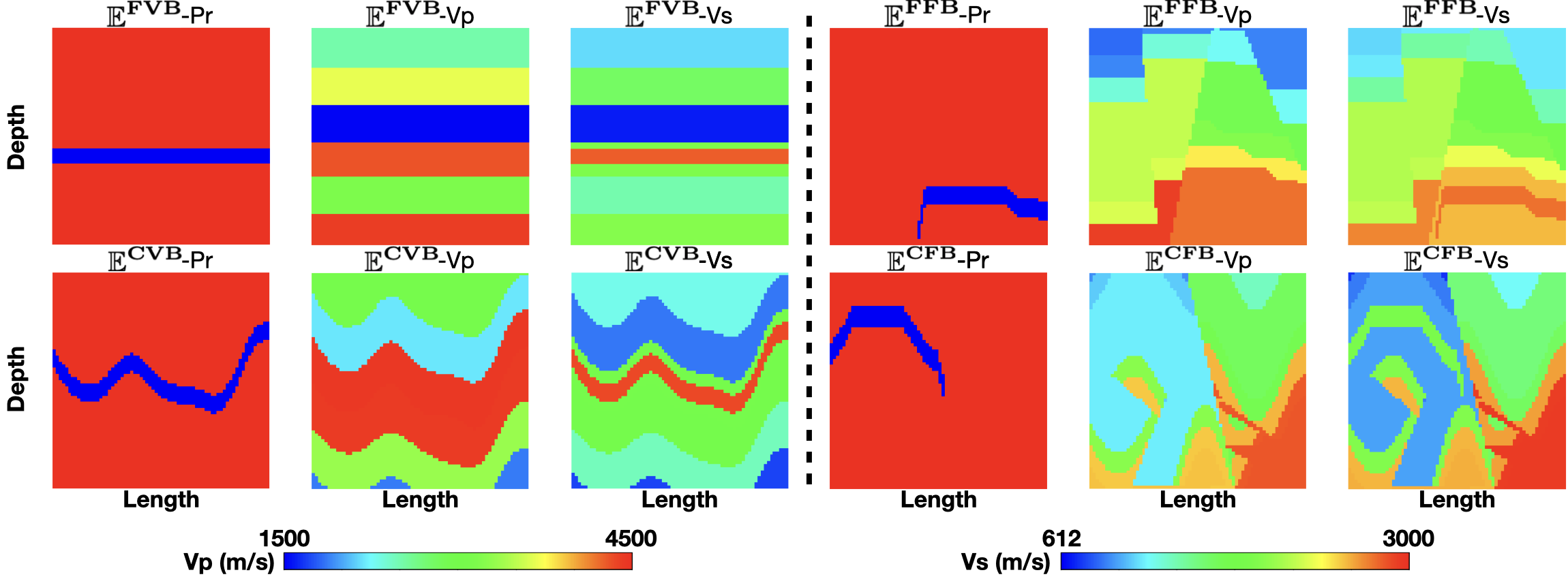}
\caption{\textbf{Gallery of $\mathbf{\mathbb{E}^{FWI}}$}: one example of reservoir structure ($\mathrm{Pr}$) and velocity maps ($V_P$, $V_S$) from the $\mathbf{\mathbb{E}^{FVB},\mathbb{E}^{FFB},\mathbb{E}^{CVB},\mathbb{E}^{CFB}}$ datasets. $\mathrm{Pr}$ refers to the designed reservoir, which is the Poisson's ratio anomaly calculated explicitly by $V_P$ and $V_S$, two kinds of wave traveling speed at each spatial point. }
\label{fig:gallery}
\end{figure*}

\section{Introduction}
\label{sec:Intro}
Seismic imaging is similar to how submarines use sonar to map out underwater landscapes and locate objects. Just as sonar sends out sound waves and interprets the echoes to figure out distances and shapes of underwater objects, geoscientists send seismic waves deep into the Earth. By analyzing how these waves are reflected back, they can generate detailed images of the subsurface and deduce properties of rock formations.

Seismic waves, propagating through the subsurface medium, can unveil the physical properties of the rock formations. Full waveform inversion (FWI) has emerged as an effective technique for obtaining high-resolution models of the subsurface physical properties~\cite{Virieux-2009-Overview,virieux2017introduction,fichtner2011resolution}. In essence, it's like refining our underwater sonar map to capture more details and nuances. The determination of such properties from seismic data is posed as an inverse problem. This means we use the reflected waves (akin to sonar echoes) to infer the properties of the rocks they passed through. FWI refines this process, striving to find the most accurate representation by minimizing the difference between observed and synthetic seismic data~\cite{fichtner2013multiscale}. This technique has made substantial contributions across a range of domains, including geothermal energy exploration, earthquake monitoring, subsurface imaging for engineering applications, and many others~\cite{vigh2011full}.


Acoustic approximations have been widely employed in wavefield simulation for FWI, resulting in a substantial reduction in computational cost~\cite{pratt1999seismic,barnes2009domain}. It assumes that the subsurface medium behaves as a fluid and focuses on simulating the kinematic aspects of compressional (P) wave propagation within the medium. However, acoustic wave propagation is an oversimplified representation of real-world scenarios, as it solely considers P-wave propagation and does not adequately model the dynamics of the wavefield~\cite{hobro2014method,raknes2017challenges}. Consequently, this oversimplification leads to suboptimal accuracy of the reconstructed medium parameters~\cite{agudo2018acoustic,fu2018multiscale,fang2020effects,feng2021mpi,sears2008elastic}.

\vspace{-0.9em}
\begin{figure*}[h!]
\centering
\includegraphics[width=0.8\columnwidth]{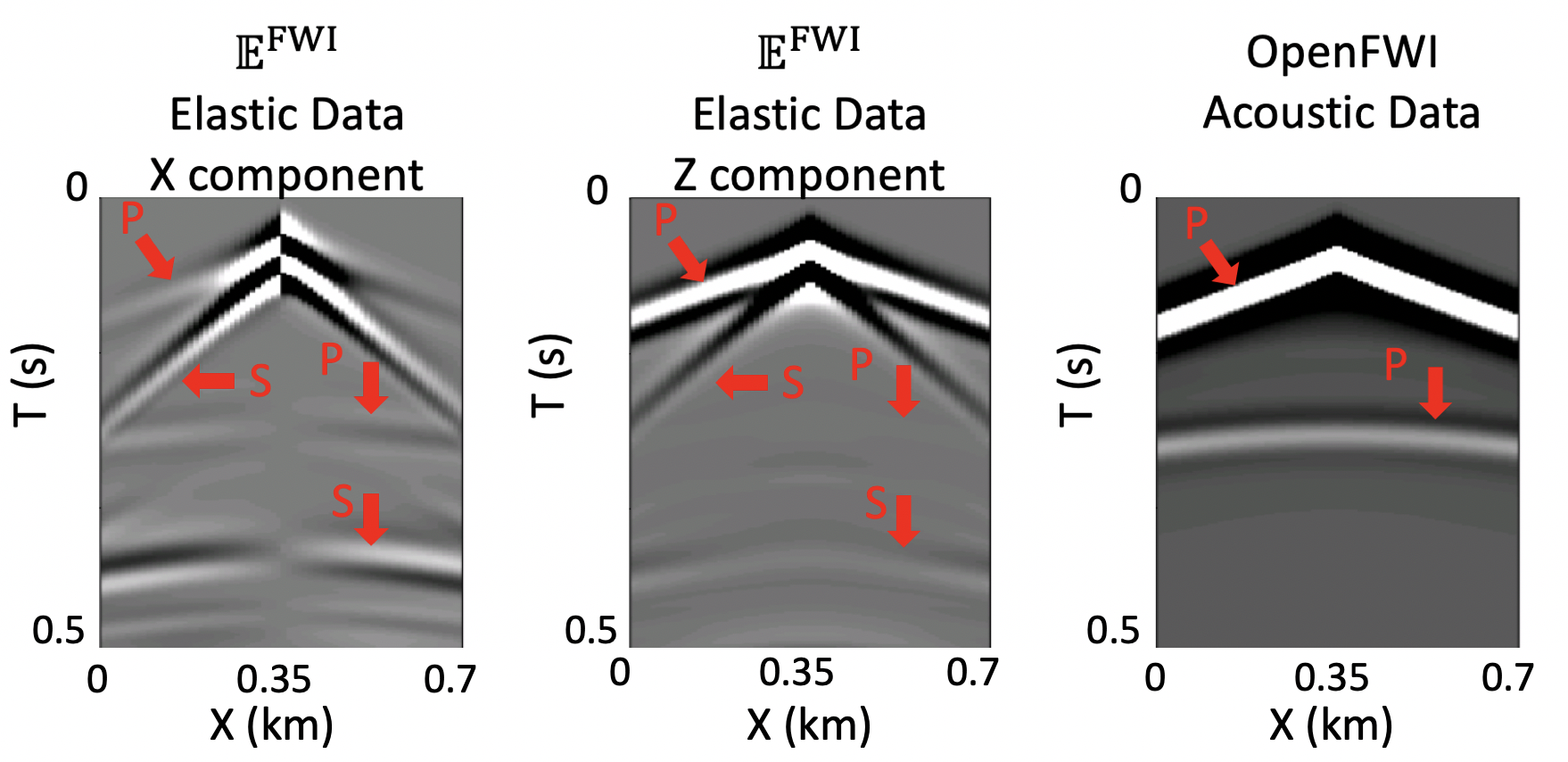}
\caption{\textbf{Comparison of elastic data in $\mathbf{\mathbb{E}^{FWI}}$  and acoustic data in \textsc{OpenFWI}}. Acoustic data only contain P-waves propagation while elastic data contain both P- and S-waves. }
\label{fig:data_compare}
\end{figure*}

\textbf{Why elastic FWI:} Elastic inversion, which considers both P- and shear (S-) waves, provides a more comprehensive and precise representation of the subsurface. The correlation between the P- (${\mathrm{V_P}}$) and S-wave velocities (${\mathrm{V_S}}$) holds significant implications in the determination of Poisson's ratio (i.e., ${\mathrm{V_P}}$-${\mathrm{V_S}}$ ratio) and Young's modulus. These parameters play a vital role in the reservoir characterization and serve as essential indicators in the identification and assessment of hydrogen and geothermal reservoirs~\cite{liu2019innovative,lees2000poisson,feng2020lithofacies,li2021elastic}. The following aspects highlight their significance:
\begin{itemize}
    \item \textit{\textbf{Lithology discrimination}: Combination of ${\mathrm{V_P}}$ and ${\mathrm{V_S}}$ velocities is useful for the lithology estimation, while ${\mathrm{V_P}}$ alone introduces significant ambiguity because of the overlap of ${\mathrm{V_P}}$ for different types of rocks~\cite{Eugenia-vpvs}.}

    \item \textit{\textbf{Fracture characterization}: Using the Poisson's ratio (${\mathrm{V_P}}$-${\mathrm{V_S}}$ ratio) and S-wave splitting can estimate fracture orientation and facilitate hydraulic fracturing stimulation~\cite{Ding-vpvs}.}

    \item \textit{\textbf{Estimation of fluid content and saturation}: Poisson's ratio (${\mathrm{V_P}}$-${\mathrm{V_S}}$ ratio) allows us to estimate the compressibility and estimate the fluid property qualitatively with other relevant reservoir parameters such as the pressure and temperature~\cite{Sun-vpvs}.}
  
\end{itemize}
 Elastic FWI, as a prominent multiparameter-inversion technique, allows us to simultaneously estimate P- and S-wave velocities~\cite{mulder2008exploring}. However, the simultaneous consideration of multiple parameters and the expanded dimensions of seismic data significantly increases the complexity of the objective function. This escalation results from the enhanced nonlinearity and the induced trade-offs between the velocities. The coupled impact of P- and S-wave velocities on seismic response further complicates the iterative update process for each parameter. Additionally, the nonlinearity becomes even more pronounced when multiple parameter classes are incorporated into the inversion, as this substantially expands the model space by introducing an increased degree of freedom~\cite{operto2013guided}. Thus, the multidimensionality of elastic FWI renders the problem considerably more complex and challenging compared to the acoustic single-parameter counterpart. With the recent development of machine learning, researchers have been actively exploring \textit{data-driven} solutions for multiparameter FWI, including multilayer perceptron (MLP)~\cite{zhang2020high}, encoder-decoder-based convolutional neural networks (CNNs)~\cite{dhara2022elastic,wu2021cnn}, recurrent network~\cite{xu2022simultaneous,zhang2020numerical}, generative adversarial networks (GANs)~\cite{yao2023regularization}, etc. Nonetheless, the absence of a publicly available elastic dataset poses challenges in facilitating a fair comparison of these methods.


To illustrate the essence of these parameters and the efficacy of elastic FWI, we spotlight the Gallery of $\mathbf{\mathbb{E}^{FWI}}$ in \Cref{fig:gallery}. This visualization showcases four sets of samples, each hailing from one of the datasets $\mathbf{\mathbb{E}^{FVB}},~\mathbf{\mathbb{E}^{FFB}},~\mathbf{\mathbb{E}^{CVB}},~\mathbf{\mathbb{E}^{CFB}}$. Each set encompasses three distinct subplots:
\begin{itemize}
    \item \textit{\textbf{P-wave Velocity Map (${\mathrm{V_P}}$):}: Demonstrating the speed at which Primary or P-waves traverse, these velocities provide insights into the composition and layering of the subsurface, such as the presence of fluids or gas.}
    \item \textit{\textbf{S-wave Velocity Map (${\mathrm{V_S}}$):}: Reflecting the pace of Secondary or S-waves, these velocities are sensitive to the rigidity and shear strength of the geological formations, offering a more detailed perspective on rock and sediment characteristics.}
    \item \textit{\textbf{Poisson's Ratio Map (${\mathrm{Pr}}$):}: Derived from the $V_P$ and $V_S$ maps, this visualization quantifies the subsurface's ability to deform under compressive stress. A higher Poisson's ratio typically indicates a more ductile material, while a lower value suggests a more brittle nature, making this measure crucial for understanding the geomechanical behavior of subsurface formations.}
\end{itemize}
$\mathbf{\mathbb{E}^{FWI}}$ is constructed upon our previously published open-access acoustic seismic dataset, known as~\textsc{OpenFWI}~\cite{deng2022openfwi}. Our approach incorporates the advantageous characteristics of \textit{multi-scale}, \textit{multi-domain}, and \textit{multi-subsurface-complexity}, inherited from the \textsc{OpenFWI} framework. Furthermore, $\mathbf{\mathbb{E}^{FWI}}$ entails the creation of S-wave velocity maps and employs the elastic wave equation in the forward modeling phase (\Cref{fig:data_compare}). The computational demands associated with conducting elastic forward modeling are substantial. Consequently, the availability of this dataset would significantly alleviate the burden on researchers. 


$\mathbf{\mathbb{E}^{FWI}}$ facilitates equitable comparisons across various methodologies using multiple datasets. In this study, we evaluate the effectiveness of three prominent methodologies derived from pre-existing networks, namely InversionNet~\cite{wu2019inversionnet}, VelocityGAN~\cite{zhang2020data}, and SimFWI~\cite{feng2023simplifying}. The objective of this evaluation is to establish a benchmark for future investigations. For comprehensive replication attempts, including the GitHub repository, pre-trained models, and associated licenses, we direct readers to the resources referenced in Section 1 of supplementary materials.

The rest of this paper is organized as follows: \Cref{sec:background} offers a comprehensive overview of the fundamental principles governing elastic FWI. \Cref{sec:dataset} presents a detailed description of the methodology employed in the construction of the dataset. \Cref{sec:benchmark} offers a succinct introduction to three deep learning methods employed for benchmarking purposes, alongside the presentation of inversion performance on each dataset. The investigation of the interdependence between P- and S-waves is conducted through ablation experiments, as outlined in \Cref{sec:ablation}. \Cref{sec:discussion} outlines the challenges faced and discusses the future implications of the dataset. Lastly, \Cref{sec:conclusion} offers conclusive remarks summarizing the key findings and contributions.
\begin{figure*}[h!]
\centering
\includegraphics[width=1\columnwidth]{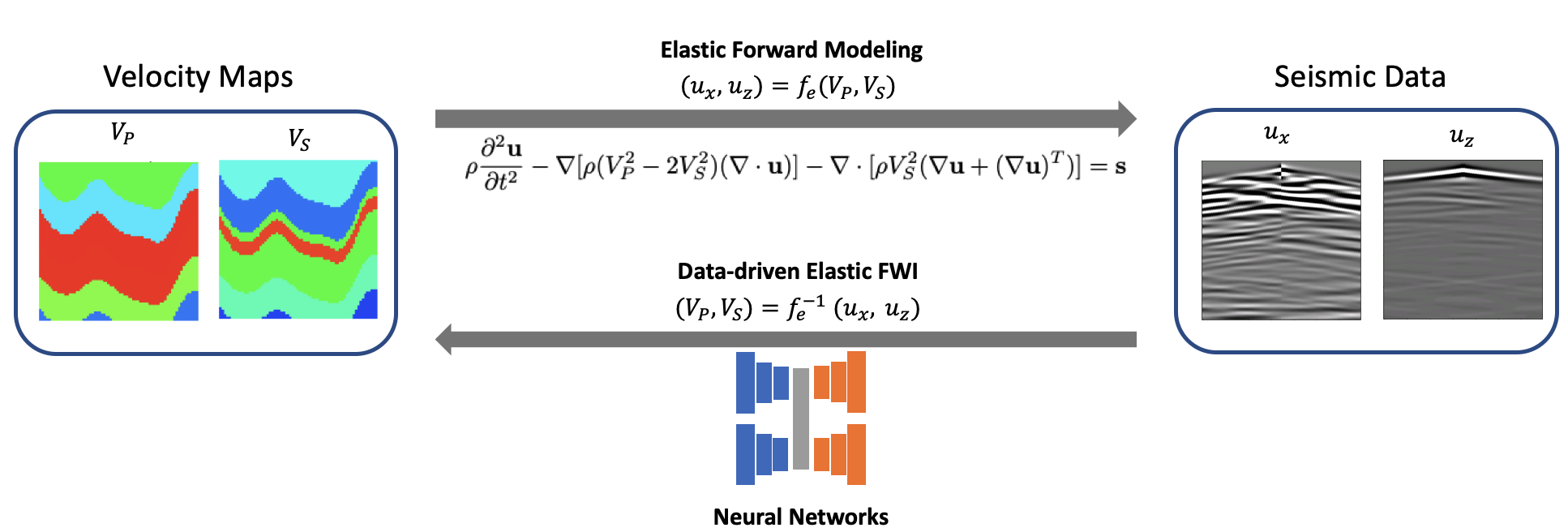}
\caption{\textbf{Schematic depiction of the data-driven approach for elastic forward modeling and FWI}. The forward modeling process involves utilizing elastic forward modeling to compute seismic data by employing the governing elastic wave equations, while elastic FWI employs neural networks to infer the P- and S-wave velocity maps from seismic data containing vertical and horizontal components.}
\label{fig:fwi_elastic}
\end{figure*}

\section{Elastic Forward Modeling and Data-driven FWI}
\label{sec:background}
\Cref{fig:fwi_elastic} provides a concise illustration of 2D data-driven elastic FWI and the relationship between P-, S-wave velocity maps and the input horizontal and vertical components of particle displacement therein. In general, the objective of data-driven elastic FWI is to employ neural networks to determine the subsurface velocity maps of the P- ($\mathrm{V_P}$) and S-waves ($\mathrm{V_S}$). The velocities depict the propagation speed of P- and S-waves through the subsurface medium and depend on the spatial coordinates $(x, z)$. We also consider the density of the subsurface as $\rho$. The source term, represented by $\mathbf{s}$, depends on spatial coordinates and time $(x, z, t)$. This source term excites both the P- and S-wave components. The particle displacement in horizontal and vertical directions is denoted by the vector $\mathbf{u} = (u_x, u_z)$. The governing equation for the elastic wave forward modeling in an isotropic medium is given as~\cite{levander1988fourth}:

\begin{equation}
   \rho\frac{\partial^2\mathbf{u}}{\partial{t}^2}-\nabla[\rho({V_P^2}-2{V_S^2})(\nabla\cdot\mathbf{u})]-\nabla\cdot[\rho{V_S^2}(\nabla\mathbf{u}+(\nabla\mathbf{u})^T)]=\mathbf{s},
   \label{eq:ewe}
\end{equation}

In the above equation:
\begin{itemize}
    \item \(\nabla\) represents the gradient calculation of a scalar field in space, specifically: 
    \[
    \nabla f = \left[\frac{\partial f}{\partial x},\frac{\partial f}{\partial z}\right]^T 
    \]
    
    \item \(\nabla \cdot\) denotes the divergence of a vector field in space, which is expressed as:
    \[
    \nabla \cdot \mathbf{v} = \frac{\partial v_x}{\partial x}+\frac{\partial v_z}{\partial z} 
    \]
\end{itemize}

For simplicity, we assume a constant density $\rho$ with the value of $1~g/cm^3$. The forward modeling problem can be expressed as $({u_x},{u_z}) = f_e({V_{P}, V_{S}})$, where $f_e(\cdot)$ signifies the highly nonlinear elastic forward modeling. It details how the P- and S-waves, initiated by the source $s$, navigate through the subsurface characterized by $\mathrm{V_P}$ and $\mathrm{V_S}$ over time $t$. Receivers then record these waves as components $u_x$ and $u_z$. The ultimate goal of data-driven elastic FWI is to harness neural networks to learn the inverse mapping $(V_{P}, V_{S}) = f_e^{-1}({u_x},{u_z})$. This inverse process enables us to deduce the subsurface velocity maps ($\mathrm{V_P}$ and $\mathrm{V_S}$) from the recorded particle displacements ($u_x$ and $u_z$) acquired from the receivers. By training neural networks with datasets of recorded waveforms and matching velocity maps, we can fine-tune the network parameters for a precise estimation of the subsurface velocities.

\section{$\mathbf{\mathbb{E}^{FWI}}$ Dataset}
\label{sec:dataset}
This section describes the methodology used to extend the velocity maps from the \textsc{OpenFWI} dataset to elastic FWI and generate our new dataset $\mathbf{\mathbb{E}^{FWI}}$. Our intention is to provide an accessible, open-source benchmark dataset that can comprehensively facilitate the development and evaluation of the machine learning algorithms in elastic FWI. 

The basic information and physical meaning of all the datasets in $\mathbf{\mathbb{E}^{FWI}}$ is summarized in~\Cref{tab:summary} and~\Cref{tab:physics}. The velocity maps encompass P- $\mathrm{V_P}$ and S-wave velocities $\mathrm{V_S}$, whereas the seismic data comprise the horizontal and vertical components of particle displacement, $u_x$ and $u_z$. The geophysical attributes in ``$\mathbf{\mathbb{E}^{Vel}}$ Family'' and the ``$\mathbf{\mathbb{E}^{Fault}}$ Family''  have been constructed utilizing the $\mathrm{V_P}$ maps derived from two distinct groups, namely the ``\textit{Vel} Family'' and the ``\textit{Fault} Family'' within the~\textsc{OpenFWI} dataset. Similar to~\textsc{OpenFWI}, the dataset has been categorized into two distinct versions, namely easy (A) and hard (B), based on the relative complexity of the subsurface structures. A thorough examination of the methodologies employed in the construction of the $\mathrm{V_P}$ maps and the detailed analysis of the complexity inherent in the velocity maps can be found in~\cite{deng2022openfwi}.

\begin{table}
\centering
\caption{\textbf{Dataset summary of $\mathbf{\mathbb{E}^{FWI}}$}. Velocity maps are represented in dimensions of depth $\times$ width $\times$ length, while seismic data is presented as \#sources $\times$ time $\times$ \#receivers in width $\times$ \#receivers in length.}
\vspace{0.5em}
\label{tab:summary}
\begin{adjustbox}{width=0.8\textwidth}
\begin{tabular}{c|c|cccc} 
\thickhline
Group                                                                                            & Dataset                       & Size    & \#Train/\#Test & Seismic Data Size                 & Velocity Map Size        \\ 
\thickhline

\multirow{2}{*}{\begin{tabular}[c]{@{}c@{}}$\mathbf{\mathbb{E}^{Vel}}$ \\ Family\end{tabular}}   & $\mathbf{\mathbb{E}^{FVA/B}}$ & $123$GB & $24$K / $6$K   & $5 \times 1000\times 1 \times 70$ & $70 \times 1 \times 70$  \\

& $\mathbf{\mathbb{E}^{CVA/B}}$ & $123$GB & $24$K / $6$K   & $5 \times 1000 \times 1\times 70$ & $70 \times 1 \times 70$  \\ 
\hline
\multirow{2}{*}{\begin{tabular}[c]{@{}c@{}}$\mathbf{\mathbb{E}^{Fault}}$ \\ Family\end{tabular}} & $\mathbf{\mathbb{E}^{FFA/B}}$ & $222$GB & $48$K / $6$K   & $5 \times 1000\times 1 \times 70$ & $70 \times 1 \times 70$  \\

& $\mathbf{\mathbb{E}^{CFA/B}}$ & $222$GB & $48$K / $6$K   & $5 \times 1000 \times 1\times 70$ & $70 \times 1 \times 70$  \\
\thickhline

\end{tabular}
\end{adjustbox}
\end{table}


The P-wave velocity ($\mathrm{V_P}$) maps in $\mathbf{\mathbb{E}^{FWI}}$ are identical to those in the previously published~\textsc{OpenFWI} dataset. For example, $\mathrm{V_P}$ in $\mathbf{\mathbb{E}^{FVA}}$ is corresponding to ``FlatVel-A'' in~\textsc{OpenFWI}, $\mathrm{V_P}$ in $\mathbf{\mathbb{E}^{CFB}}$ is corresponding to ``CurveFault-B'' in~\textsc{OpenFWI}, and the same naming rule applies to the rest datasets. These velocity maps incorporate a wide range of geological scenarios reflecting diverse subsurface complexities, thereby providing an extensive testbed for machine learning methodologies.

In order to construct the S-wave velocity ($\mathrm{V_S}$) maps, we incorporate the Poisson's ratio ($\mathrm{Pr}$) maps~\cite{christensen1996poisson}, which provide a representation of the relationship between the P- ($\mathrm{V_P}$) and the S-wave velocities ($\mathrm{V_S}$)
\begin{eqnarray}
P_r&=&\frac{V^2_P-2V^2_S}{2V^2_P-2V^2_S}.
\label{Possion}
\end{eqnarray}
The initial step involves the generation of Poisson's ratio ($\mathrm{Pr}$) maps by selecting two values within the reasonable range of 0.1 to 0.4 in a random manner~\cite{gercek2007poisson}. One of these values is allocated to represent the background, whereas the other value is assigned to represent a thin-layer reservoir. Thin-layer reservoirs are selected due to their significance in representing areas where pores are saturated with fluids, making them crucial targets for subsurface exploration and reservoir detection. In the $\mathbf{\mathbb{E}^{FWI}}$ framework, the S-wave velocity ($\mathrm{V_S}$) maps are synthesized by multiplying the models of P-wave velocity ($\mathrm{V_P}$) with the respective Poisson's ratio ($\mathrm{Pr}$) maps, adhering to the following relationship:
\begin{equation}
    V_S = \sqrt{\frac{0.5 - Pr}{1- Pr}}*V_P.
\end{equation}

 This approach ensures a wide range of velocity contrasts, resulting in diverse wavefield behaviors, thus expanding the scope of scenarios for machine learning tests in elastic FWI. The details of the elastic forward modeling are given in Section 2 of the supplementary materials.


\begin{table}
\centering
\caption{\textbf{Physical meaning of $\mathbf{\mathbb{E}^{FWI}}$ dataset}}
\vspace{0.5em}
\renewcommand{\arraystretch}{1.5}
\label{tab:physics}
\begin{adjustbox}{width=1\textwidth}
\begin{tabular}{c|c|c|c|c|c|c|c|c} 
\thickhline
Dataset                & \begin{tabular}[c]{@{}c@{}}Grid\\Spacing\end{tabular} & \begin{tabular}[c]{@{}c@{}}Velocity Map\\Spatial Size\end{tabular} & \begin{tabular}[c]{@{}c@{}}Source\\~Spacing\end{tabular} & \begin{tabular}[c]{@{}c@{}}Source Line\\Length\end{tabular} & \begin{tabular}[c]{@{}c@{}}Receiver Line\\Spacing\end{tabular} & \begin{tabular}[c]{@{}c@{}}Receiver Line\\Length\end{tabular} & \begin{tabular}[c]{@{}c@{}}Time\\~Spacing\end{tabular} &  \begin{tabular}[c]{@{}c@{}}Recorded\\Time\end{tabular}  \\ 
\thickhline
$\mathbf{\mathbb{E}^{Vel}}$, $\mathbf{\mathbb{E}^{Fault}}$ Family & 5 $m$                                                & 0.35 $\times$ 0.35 $km^2$                                           & 87.5 $m$                                                  & 0.35 $km$                                                    & 5 $m$                                                    & 0.35 $km$                                                 & 0.001 $s$                                                                                            & 1 $s$                                                   \\
\thickhline
\end{tabular}
\end{adjustbox}
\end{table}

\section{$\mathbf{\mathbb{E}^{FWI}}$ Benchmarks}
\label{sec:benchmark}
\subsection{Deep Learning Methods for Elastic FWI}
Our benchmark presents inversion results by three deep learning-based approaches, namely $\mathbf{\mathbb{E}}$lasticNet, $\mathbf{\mathbb{E}}$lasticGAN, and $\mathbf{\mathbb{E}}$lasticTransformer. These methods are derived from pre-existing networks, namely InversionNet~\cite{wu2019inversionnet}, VelocityGAN~\cite{zhang2020data}, and SimFWI~\cite{feng2023simplifying}, with modifications tailored to address the challenges posed by elastic FWI.
We provide a summary of each method separately as follows:

\textbf{$\mathbf{\mathbb{E}}$lasticNet} is extended from the vanilla InversionNet~\cite{wu2019inversionnet} to the elastic setting with two pairs of input and output. It is a fully-convolutional neural network taking seismic data $u_x$ and $u_z$ as 
the input of two encoders to learn the latent embeddings independently. The mutual representations of two inputs are concatenated and then forwarded to two independent decoders to obtain the estimated velocity maps $\mathrm{V_P}$ and $\mathrm{V_S}$ as output. 

\textbf{$\mathbf{\mathbb{E}}$lasticGAN} follows the design of VelocityGAN~\cite{zhang2020data} but substitutes the original generator with an encoder-decoder network such as $\mathbf{\mathbb{E}}$lasticNet. The estimated velocity maps $\mathrm{V_P}$ and $\mathrm{V_S}$ produced by the generator are fed to two independent discriminators to identify the real and fake predictions. A CNN architecture is employed for both discriminators.

\textbf{$\mathbf{\mathbb{E}}$lasticTransformer} follows a similar seismic-encoder and velocity-decoder architecture design as the SimFWI described in \cite{feng2023simplifying}. 
It consists of two two-layer transformer encoders that take $u_x$ and $u_z$ as inputs and two two-layer transformer decoders to output $\mathrm{V_P}$ and $\mathrm{V_S}$ separately. Two latent embeddings of $u_x$ and $u_z$ are concatenated and passed through two Maxout converters, then transformed embeddings fed into the decoders.
Unlike the linear upsampler utilized at the end of the velocity decoder in \cite{feng2023simplifying}, we stack upsampling and convolution blocks to construct the upsampler.

\subsection{Inversion Benchmarks}

\begin{figure*}[h!]
\centering
\includegraphics[width=1.0\columnwidth]{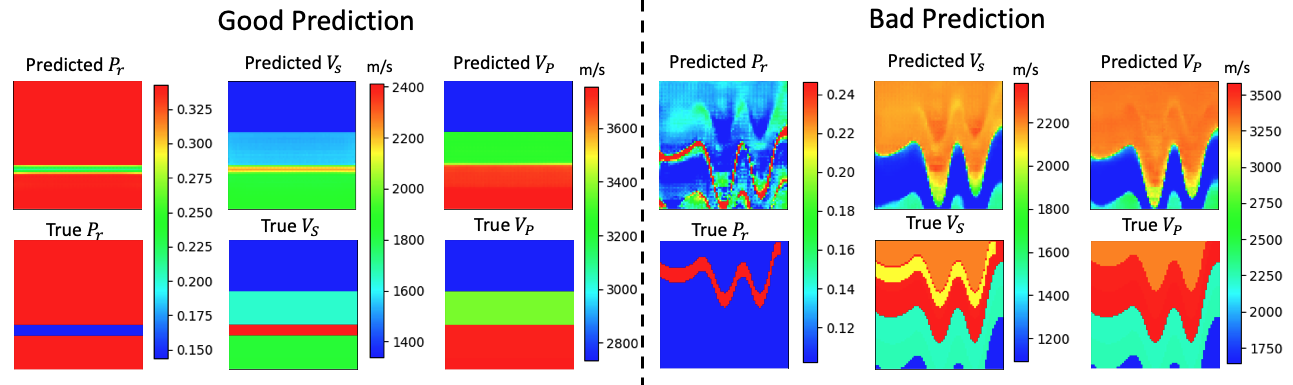}
    \caption{\textbf{Examples of both successful and inadequate predictions} in $\mathbf{\mathbb{E}}^{FWI}$ benchmarks performed by the $\mathbf{\mathbb{E}}$lasticNet.}
\label{fig:prediction_example}
\end{figure*}

The experiments were conducted using Nvidia Tesla V100 GPUs, and the training parameters were kept consistent across all datasets. The training process is conducted utilizing the $\ell_1$-norm and $\ell_2$-norm loss functions, respectively. In our study, we assess not only the accuracy of the predicted velocities $\mathrm{V_P}$ and $\mathrm{V_S}$ but also the degree of decoupling between them by evaluating the accuracy of the predicted Poisson ratio $(\mathrm{Pr})$. To quantify the performance of our predictions, we utilize three evaluation metrics: mean absolute error (MAE), root mean square error (RMSE), and structural similarity index (SSIM). These metrics provide a comprehensive assessment of the quality of our predictions and their similarity to the ground truth values. The performance of $\mathbb{E}$lasticNet on various datasets is presented in \Cref{table:InversionNet_Benchmark2D}, while \Cref{tab:time} provides the estimated training time per epoch for each method on the $\mathbb{E}^{FWI}$ datasets. In \Cref{fig:prediction_example}, examples of inverted velocity maps obtained using the $\mathbb{E}$lasticNet are presented alongside the corresponding ground truth velocity maps. These visual representations highlight instances where the inversion process successfully predicts accurate velocities, as well as instances where further improvement is required. The benchmarks with ${\mathbb{E}}$lasticGAN and {${\mathbb{E}}$lasticTransformer} are given in Section 6 of the supplementary materials.

The performance of all three models exhibits a decline as the complexity of the dataset increases. Notably, in the case of the dataset with version B, it consistently exhibits lower performance compared to the dataset with version A. The network provides direct predictions for $\mathrm{V_P}$ and $\mathrm{V_S}$, whereas $\mathrm{Pr}$ is obtained indirectly through calculations based on $\mathrm{V_P}$ and $\mathrm{V_S}$. As a result, $\mathrm{Pr}$ consistently exhibits lower SSIM compared to $\mathrm{V_P}$ and $\mathrm{V_S}$. However, it should be noted that $\mathrm{Pr}$ represents a sparser map compared to $\mathrm{V_P}$ and $\mathrm{V_S}$, leading to lower MAE and RMSE values for $\mathrm{Pr}$ compared to $\mathrm{V_P}$ and $\mathrm{V_S}$.

\vspace*{-5mm}
\begin{table}[!ht]
\footnotesize
\renewcommand{\arraystretch}{1.3}
\centering
\caption{\textbf{Quantitative results} of ${\mathbb{E}}$lasticNet on $\mathbf{\mathbb{E}^{FWI}}$ datasets. }
\vspace{0.5em}
\begin{adjustbox}{width=1\textwidth}
\begin{tabular}{c|cc|ccc|ccc|ccc} 
\thickhline
\multirow{3}{*}{Dataset}      & \multicolumn{2}{c|}{\multirow{3}{*}{Loss}} & \multicolumn{9}{c}{${\mathbb{E}}$lasticNet}    \\ 
\cline{4-12}
    & \multicolumn{2}{c|}{}
    & \multicolumn{3}{c|}{$\mathrm{V_P}$}
    & \multicolumn{3}{c|}{$\mathrm{V_S}$}
    & \multicolumn{3}{c}{$\mathrm{Pr}$}
    \\ 
\cline{4-12}
    & \multicolumn{2}{c|}{}
    & \multicolumn{1}{c}{MAE$\downarrow$}    & RMSE$\downarrow$    & SSIM$\uparrow$  
    & \multicolumn{1}{c}{MAE$\downarrow$}    & RMSE$\downarrow$    & SSIM$\uparrow$    
    & \multicolumn{1}{c}{MAE$\downarrow$}    & RMSE$\downarrow$    & SSIM$\uparrow$    \\ 
\hline
\multirow{2}{*}{$\mathbf{\mathbb{E}^{FVA}}$}
    & $\ell_1$ &
    & 0.0308    & 0.0559    & 0.9615
    & 0.0259    & 0.0500    & 0.9596    
    & 0.0329    & 0.0664    & 0.8455    \\
    & & $\ell_2$
    & 0.0235    & 0.0455    & \textbf{0.9702}
    & 0.0196    & 0.0385    & \textbf{0.9683}
    & 0.0307    & 0.0583    & \textbf{0.8644}    \\ 
\hline
\multirow{2}{*}{$\mathbf{\mathbb{E}^{FVB}}$}
    & $\ell_1$ &
    & 0.0668    & 0.1468    & \textbf{0.8891}
    & 0.0483    & 0.1053    & \textbf{0.8951}
    & 0.0542    & 0.1057    & \textbf{0.7138}    \\
    & & $\ell_2$
    & 0.1016    & 0.1901    & 0.8354
    & 0.0691    & 0.1322    & 0.8599
    & 0.0756    & 0.1302    & 0.6227    \\ 
\hline
\multirow{2}{*}{$\mathbf{\mathbb{E}^{CVA}}$}
    & $\ell_1$ &
    & 0.0745    & 0.1345    & 0.8055
    & 0.0600    & 0.1080    & 0.8051
    & 0.0574    & 0.1156    & 0.5766    \\
    & & $\ell_2$    
    & 0.0745    & 0.1343    & \textbf{0.8033} 
    & 0.0616    & 0.1087    & \textbf{0.8020} 
    & 0.0604    & 0.1131    & \textbf{0.6131}    \\ 
\hline
\multirow{2}{*}{$\mathbf{\mathbb{E}^{CVB}}$}
    & $\ell_1$ &
    & 0.1722    & 0.2982    & 0.6529
    & 0.1258    & 0.2165    & 0.6827 
    & 0.0915    & 0.1580    & \textbf{0.4612}    \\
    & & $\ell_2$
    & 0.1682    & 0.3048    & \textbf{0.6566}
    & 0.1234    & 0.2220    & \textbf{0.6875}
    & 0.0956    & 0.1660    & 0.4337    \\ 
\hline
\multirow{2}{*}{$\mathbf{\mathbb{E}^{FFA}}$}
    & $\ell_1$ &
    & 0.0543    & 0.1026    & \textbf{0.9042}
    & 0.0647    & 0.1349    & 0.8225
    & 0.0710    & 0.1501    & \textbf{0.6447}    \\
    & & $\ell_2$    
    & 0.0937    & 0.1537    & 0.8607
    & 0.0769    & 0.1309    & \textbf{0.8305}
    & 0.0830    & 0.1369    & 0.6251    \\ 
\hline
\multirow{2}{*}{$\mathbf{\mathbb{E}^{FFB}}$}
    & $\ell_1$ &    
    & 0.1198    & 0.1859    & 0.7014
    & 0.0947    & 0.1462    & 0.7346
    & 0.0802    & 0.1312    & 0.4902    \\
    & & $\ell_2$
    & 0.1084    & 0.1704    & \textbf{0.7131}
    & 0.0811    & 0.1290    & \textbf{0.7523}
    & 0.0719    & 0.1225    & \textbf{0.5270}    \\ 
\hline
\multirow{2}{*}{$\mathbf{\mathbb{E}^{CFA}}$} 
    & $\ell_1$ &
    & 0.0551    & 0.1128    & \textbf{0.8814}
    & 0.0518    & 0.1042    & \textbf{0.8445}
    & 0.0528    & 0.1150    & \textbf{0.6562}    \\
    & & $\ell_2$   
    & 0.0972    & 0.1636    & 0.8223
    & 0.0886    & 0.1390    & 0.7891
    & 0.0927    & 0.1443    & 0.5536    \\ 
\hline
\multirow{2}{*}{$\mathbf{\mathbb{E}^{CFB}}$} 
    & $\ell_1$ &
    & 0.1535    & 0.2307    & 0.5981 
    & 0.1123    & 0.1698    & 0.6408
    & 0.1012    & 0.1602    & 0.3576    \\
    & & $\ell_2$
    & 0.1562    & 0.2305    & \textbf{0.6160}    
    & 0.1138    & 0.1697    & \textbf{0.6608} 
    & 0.0854    & 0.1393    & \textbf{0.4490}    \\
\thickhline
\end{tabular}
\end{adjustbox}
\vspace{0.5em}
\label{table:InversionNet_Benchmark2D}
\vspace{-1em}
\end{table}

\begin{table}
\centering
\caption{\textbf{Training time} in each epoch by each benchmarking method on $\mathbf{\mathbb{E}^{FWI}}$  datasets. All the models are trained on a single GPU.}
    \vspace{0.5em}
\begin{adjustbox}{width=0.7\textwidth}
\label{tab:time}
\begin{tabular}{c|ccc} 
\thickhline
             & $\mathbf{\mathbb{E}}$lasticNet & $\mathbf{\mathbb{E}}$lasticGAN & $\mathbf{\mathbb{E}}$lasticTransformer   \\ 
\hline
$\mathbf{\mathbb{E}^{Vel}}$ Family & 4m15s         & 2m20s           & 1m15s                    \\
$\mathbf{\mathbb{E}^{Fault}}$ Family& 8m35s           & 3m50s             &  2m30s                      \\
\thickhline
\end{tabular}
\end{adjustbox}
\end{table}

\section{Ablation Study}
\label{sec:ablation}
\subsection{Independent vs. Joint Inversion: Impact on $\mathbf{\mathrm{Pr}}$ Maps}
The first experiment examined the impact of separate versus joint inversion of $\mathrm{V_P}$ and $\mathrm{V_S}$ on the accuracy of predicted Poisson ratio ($\mathrm{Pr}$) maps. This process involved individually training two InversionNets on the $\mathbf{\mathbb{E}^{FWI}}$ dataset to predict $\mathrm{V_P}$ and $\mathrm{V_S}$ maps, which were then used to calculate the $\mathrm{Pr}$ maps. The results revealed a substantial deterioration in map quality, with the independent inversion maps exhibiting significantly higher MAE and MSE, and lower SSIM values, as outlined in~\Cref{table:thin_layer_vp},  compared to those reconstructed from joint inversion, shown in~\Cref{table:InversionNet_Benchmark2D}, especially for the complex B datasets, such as ``$\mathbf{\mathbb{E}^{CFB}}$''. These findings reinforce the significance of considering the $\mathrm{V_P}$-$\mathrm{V_S}$ relationship and P-S wave coupling, with the single-parameter inversion approach being deemed unviable. Detailed information on this experiment can be found in Section 6 of the supplementary materials.

\subsection{Investigating P- and S-waves Coupling via Machine Learning}

The second experiment focused on examining the interaction between P- and S-waves in the context of seismic data inversion. Two InversionNets were trained, one focusing on P-wave velocity ($\mathrm{V_P}$) and the other on S-wave velocity ($\mathrm{V_S}$), while adjusting the structural characteristics of the ignored wave. This experiment, trained with the \textsc{OpenFWI}'s InversionNet using data from $\mathbf{\mathbb{E}^{FWI}}$, revealed that any minor change in the disregarded wave velocity structure significantly degraded the network's performance, as evidenced in~\Cref{table:change_structure}. This outcome was clearly demonstrated in the more complex datasets, such as ``$\mathbf{\mathbb{E}^{CFB}}$'' test set, where changes in structure led to a substantial increase in MAE and RMSE, along with a decrease in the SSIM. For a more detailed analysis, refer to the supplementary materials.

\begin{table}
\footnotesize
\renewcommand{\arraystretch}{1.3}
\centering
\caption{\textbf{Quantitative results} of \textsc{OpenFWI}'s InversionNet trained with $\mathbb{E}^{FWI}$ data. Input z-component data and output the $\mathrm{V_P}$/$\mathrm{V_S}$ maps independently. The performance is a benchmark of $\mathbb{E}$lasticNet. }
\vspace{0.5em}
\begin{adjustbox}{width=1\textwidth}
\begin{tabular}{c|cc|ccc|ccc|ccc} 
\thickhline
\multirow{3}{*}{Dataset}      & \multicolumn{2}{c|}{\multirow{3}{*}{Loss}} & \multicolumn{9}{c}{InversionNet}    \\ 
\cline{4-12}
    & \multicolumn{2}{c|}{}    
    & \multicolumn{3}{c|}{$\mathrm{V_P}$}
    & \multicolumn{3}{c|}{$\mathrm{V_S}$} 
    & \multicolumn{3}{c} {Pr} \\ 
\cline{4-12}
    & \multicolumn{2}{c|}{}
    & \multicolumn{1}{c}{MAE$\downarrow$} & RMSE$\downarrow$ & SSIM$\uparrow$  
    & \multicolumn{1}{c}{MAE$\downarrow$} & RMSE$\downarrow$ & SSIM$\uparrow$
    & \multicolumn{1}{c}{MAE$\downarrow$} & RMSE$\downarrow$ & SSIM$\uparrow$    \\ 
\hline
\multirow{2}{*}{$\mathbf{\mathbb{E}^{FVA}}$}    
    & $\ell_1$ &
    & 0.0392    & 0.0712    & \textbf{0.9455}
    & 0.0239    & 0.0447    & \textbf{0.9590}
    & 0.0461    & 0.0885    & \textbf{0.8282}    \\
    & & $\ell_2$    
    & 0.0451    & 0.0745    & 0.9414
    & 0.0251    & 0.0469    & 0.9585
    & 0.0541    & 0.1039    & 0.8071    \\ 
\hline
\multirow{2}{*}{$\mathbf{\mathbb{E}^{FVB}}$}
    & $\ell_1$ &
    & 0.1030    & 0.1986    & 0.8260 
    & 0.0643    & 0.1318    & 0.8615 
    & 0.1290    & 0.2773    & 0.6063    \\
    &  & $\ell_2$
    & 0.0883    & 0.1832    & \textbf{0.8453}
    & 0.0620    & 0.1269    & \textbf{0.8665}
    & 0.0905    & 0.1980    & \textbf{0.6603}    \\
\hline
\multirow{2}{*}{$\mathbf{\mathbb{E}^{CVA}}$}
    & $\ell_1$ &
    & 0.1016    & 0.1699    & \textbf{0.7636}    
    & 0.0736    & 0.1245    & \textbf{0.7837}    
    & 0.1050    & 0.2141    & \textbf{0.5607}    \\
    &  & $\ell_2$
    & 0.1052    & 0.1730    & 0.7460
    & 0.0720    & 0.1236    & 0.7798
    & 0.1194    & 0.2381    & 0.5454    \\ 
\hline
\multirow{2}{*}{$\mathbf{\mathbb{E}^{CVB}}$}
    & $\ell_1$ &
    & 0.1854    & 0.3266    & 0.6270 
    & 0.1388    & 0.2405    & 0.6578 
    & 0.1523    & 0.3064    & \textbf{0.4695}    \\
    &  & $\ell_2$
    & 0.1820    & 0.3260    & \textbf{0.6323}
    & 0.1331    & 0.2344    & \textbf{0.6674}
    & 0.1553    & 0.3230    & 0.4648    \\ 
\hline
\multirow{2}{*}{$\mathbf{\mathbb{E}^{FFA}}$}
    & $\ell_1$ &
    & 0.0818    & 0.1413    & 0.8625
    & 0.0584    & 0.1034    & \textbf{0.8681}
    & 0.0784    & 0.1585    & \textbf{0.6937}    \\
    &  & $\ell_2$
    & 0.0788    & 0.1343    & \textbf{0.8930} 
    & 0.0946    & 0.1525    & 0.7918
    & 0.1351    & 0.2485    & 0.6174    \\ 
\hline
\multirow{2}{*}{$\mathbf{\mathbb{E}^{FFB}}$}
    & $\ell_1$ &    
    & 0.1323    & 0.2001    & 0.6790 
    & 0.0943    & 0.1453    & 0.7312
    & 0.1180    & 0.2283    & 0.5280    \\
    &  & $\ell_2$
    & 0.1301    & 0.1979    & \textbf{0.6808}
    & 0.0898    & 0.1382    & \textbf{0.7399}
    & 0.1124    & 0.2095    & \textbf{0.5494}    \\ 
\hline
\multirow{2}{*}{$\mathbf{\mathbb{E}^{CFA}}$}
    & $\ell_1$ &
    & 0.1012    & 0.1638    & 0.8624
    & 0.0710    & 0.1182    & \textbf{0.8485}
    & 0.0778    & 0.1586    & \textbf{0.6857}    \\
    &  & $\ell_2$    
    & 0.0962    & 0.1663    & \textbf{0.8443}
    & 0.0833    & 0.1394    & 0.8106
    & 0.1032    & 0.1911    & 0.6393    \\ 
\hline
\multirow{2}{*}{$\mathbf{\mathbb{E}^{CFB}}$}
    & $\ell_1$ &    
    & 0.1702    & 0.2485    & \textbf{0.6020}
    & 0.1243    & 0.1807    & 0.6531
    & 0.1317    & 0.2378    & \textbf{0.5091}    \\
    &  & $\ell_2$
    & 0.1745    & 0.2563    & 0.5849
    & 0.1219    & 0.1775    & \textbf{0.6588}
    & 0.1379    & 0.2529    & 0.4841    \\
\thickhline
\end{tabular}
\end{adjustbox}
\vspace{0.5em}
\label{table:thin_layer_vp}
\vspace{-1em}
\end{table}

\begin{table}
\footnotesize
\renewcommand{\arraystretch}{1.3}
\centering
\caption{\textbf{Quantitative results} of InversionNet trained with $\mathbf{\mathbb{E}^{FWI}}$ data. The performance compared between testing on the same and different disregarded velocity structural datasets. }
\vspace{0.5em}
\begin{adjustbox}{width=0.75\textwidth}
\begin{tabular}{c|cc|ccc|ccc} 
\thickhline
\multirow{3}{*}{Dataset}      & \multicolumn{2}{c|}{\multirow{3}{*}{Loss}} & \multicolumn{6}{c}{InversionNet}    \\ 
\cline{4-9}
    & \multicolumn{2}{c|}{}    
    & \multicolumn{3}{c|}{$\mathrm{V_P}$: different $\mathrm{V_S}$ structure}
    & \multicolumn{3}{c} {$\mathrm{V_S}$: different $\mathrm{V_P}$ structure} \\ 
\cline{4-9}
    & \multicolumn{2}{c|}{}
    & \multicolumn{1}{c}{MAE$\downarrow$} & RMSE$\downarrow$ & SSIM$\uparrow$  
    & \multicolumn{1}{c}{MAE$\downarrow$} & RMSE$\downarrow$ & SSIM$\uparrow$    \\ 
\hline
\multirow{2}{*}{$\mathbf{\mathbb{E}^{FVA}}$}    
    & $\ell_1$ &
    & 0.1162    & 0.1919    & \textbf{0.8974} 
    & 0.2040    & 0.2716    & \textbf{0.7977}   \\
    & & $\ell_2$    
    & 0.1245    & 0.2101    & 0.8849
    & 0.2189    & 0.2928    & 0.7793  \\ 
\hline
\multirow{2}{*}{$\mathbf{\mathbb{E}^{FVB}}$}
    & $\ell_1$ &
    & 0.2098    & 0.3527    & 0.7086   
    & 0.2479    & 0.3285    & \textbf{0.7182}    \\
    &  & $\ell_2$
    & 0.1940    & 0.3265    & \textbf{0.7279}    
    & 0.2540    & 0.3382    & 0.7022  \\
\hline
\multirow{2}{*}{$\mathbf{\mathbb{E}^{CVA}}$}
    & $\ell_1$ &
    & 0.1624    & 0.2590    & \textbf{0.7233}
    & 0.2202    & 0.2924    & 0.6794   \\
    &  & $\ell_2$
    & 0.1678    & 0.2647    & 0.7043
    & 0.2249    & 0.2987    & \textbf{0.6824}   \\ 
\hline
\multirow{2}{*}{$\mathbf{\mathbb{E}^{CVB}}$}
    & $\ell_1$ &
    & 0.3014    & 0.4828    & \textbf{0.5206}
    & 0.3136    & 0.4246    & \textbf{0.5398}  \\
    &  & $\ell_2$
    & 0.3067    & 0.4900    & 0.5181
    & 0.3207    & 0.4349    & 0.5341  \\ 
\hline
\multirow{2}{*}{$\mathbf{\mathbb{E}^{FFA}}$}
    & $\ell_1$ &
    & 0.1741    & 0.2732    & 0.7464
    & 0.2200    & 0.2865    & \textbf{0.7362}     \\
    &  & $\ell_2$
    & 0.1303    & 0.2052    & \textbf{0.8490}
    & 0.2228    & 0.2913    & 0.6952    \\ 
\hline
\multirow{2}{*}{$\mathbf{\mathbb{E}^{FFB}}$}
    & $\ell_1$ &    
    & 0.1575    & 0.2291    & \textbf{0.6501}
    & 0.2350    & 0.3030    & 0.6480    \\
    &  & $\ell_2$
    & 0.1627    & 0.2336    & 0.6492
    & 0.2319    & 0.2982    & \textbf{0.6517}   \\ 
\hline
\multirow{2}{*}{$\mathbf{\mathbb{E}^{CFA}}$}
    & $\ell_1$ &
    & 0.1404    & 0.2244    & 0.7951
    & 0.2327    & 0.3059    & \textbf{0.7360}   \\
    &  & $\ell_2$    
    & 0.1319    & 0.2134    & \textbf{0.8159}
    & 0.2399    & 0.3194    & 0.7139    \\ 
\hline
\multirow{2}{*}{$\mathbf{\mathbb{E}^{CFB}}$}
    & $\ell_1$ &    
    & 0.1937    & 0.2798    & \textbf{0.5747}
    & 0.2465    & 0.3215    & 0.5785 \\
    &  & $\ell_2$
    & 0.1928    & 0.2788    & 0.5613
    & 0.2445    & 0.3183    & \textbf{0.5915}    \\
\thickhline
\end{tabular}
\end{adjustbox}
\vspace{0.5em}
\label{table:change_structure}
\vspace{-1em}
\end{table}
\section{Discussion}
\label{sec:discussion}

\subsection{Future Challenge}

\textbf{Decouple P- and S-waves:} The interaction between P- and S-waves during seismic wave propagation poses a significant challenge when attempting to simultaneously determine P and S velocities. The networks described in this paper exhibit limited success in separating P- and S-waves within the seismic data. Consequently, we anticipate the development of robust methodologies that can precisely estimate both P and S velocities while effectively mitigating the interdependence between these wave components.

\textbf{Generalization of data-driven methods:} The elastic approximation provides a more accurate representation of field data in comparison to acoustic data. As a result, we expect the neural networks trained on the $\mathbf{\mathbb{E}^{FWI}}$ dataset to exhibit improved resilience in handling real-world field data. However, it should be noted that there are additional physical phenomena, such as anisotropy and viscosity, which are not accounted for in the $\mathbf{\mathbb{E}^{FWI}}$ dataset. The question of how to incorporate these phenomena into the analysis of field data remains an open and unanswered challenge.

\textbf{Forward modeling:} The computational expense associated with elastic forward modeling surpasses that of acoustic cases due to various factors. These include the increased memory requirements and the implementation of smaller grid sizes to counteract dispersion phenomena, among others. A detailed comparison highlighting these aspects can be found in the last section of supplementary materials. Despite the possibility of bypassing extensive forward modeling by providing the $\mathbf{\mathbb{E}^{FWI}}$ dataset, there remains a need to explore an efficient forward modeling algorithm to accommodate the growing volume of data in the field.


\subsection{Broader Impact} 

\textbf{Multiparameter inversion:} Multiparameter inversion techniques have found wide-ranging applications across diverse scientific and engineering domains, including but not limited to geophysics, medical imaging, and material science. The introduction of $\mathbf{\mathbb{E}^{FWI}}$ serves as a catalyst for further investigation and the pursuit of innovative methodologies in these fields. By addressing the inherent limitations and complexities associated with multiparameter inversion, this advancement encourages ongoing research and the exploration of novel solutions.

 \textbf{Carbon-zero emission: } 
The attainment of carbon-zero emissions holds paramount significance in addressing climate change, safeguarding human well-being, and fostering sustainable development. While researchers continue to explore effective strategies towards achieving this goal, elastic FWI emerges as a promising approach that can contribute significantly. Particularly, elastic FWI plays a crucial role in assessing and developing geothermal energy resources, as well as in facilitating carbon capture and storage projects, among other applications. The introduction of $\mathbf{\mathbb{E}^{FWI}}$ as a fundamental dataset for elastic FWI is expected to stimulate further research and innovation in this direction, thereby enhancing our understanding and capabilities in the pursuit of carbon-zero emissions.

\textbf{New energy exploration:} Elastic FWI can be utilized to evaluate the geological viability of potential sites for hydrogen storage, including underground formations or depleted oil and gas reservoirs. The suitability, capacity, and feasibility of a storage site heavily rely on the effectiveness of a geophysical survey and characterization approaches. With the availability of the $\mathbf{\mathbb{E}^{FWI}}$ dataset, it would yield great potential to enhance the accuracy in  characterizing the subsurface reservoir, therefore providing better identification of hydrogen storage locations.

\textbf{Potential Social Impacts:} Our research, while advancing inverse problems in natural science, may inadvertently support increased fossil fuel consumption if applied to optimize oil and gas drilling, raising environmental concerns. However, the same techniques can equally bolster positive initiatives, such as geothermal exploration and hydrogen storage, as highlighted in Sec 6.2. The dual potential of our methodologies underscores the importance of their judicious application, aligning with sustainable development goals.


\section{Conclusion}
\label{sec:conclusion}
This paper presents $\mathbf{\mathbb{E}^{FWI}}$, an open-source elastic FWI dataset.~$\mathbf{\mathbb{E}^{FWI}}$ comprises eight datasets and includes benchmarks for three deep learning methods. The datasets released with~$\mathbf{\mathbb{E}^{FWI}}$, provide diverse P-wave and S-wave velocities, specifically addressing the coupling problem encountered in multiparameter inversion. The initial benchmarks demonstrate promising results on certain datasets, while others may require further investigation. Additionally, coupling tests are conducted to provide insights into network design for multiparameter inversion problems. Furthermore, this paper discusses the future challenges that can be explored using these datasets and outlines the envisioned future advancements as $\mathbf{\mathbb{E}^{FWI}}$ continues to evolve.


\section*{Acknowledgement}

This work was funded by the Los Alamos National Laboratory~(LANL) - Technology Evaluation and Demonstration~(TED) program and by the U.S. Department of Energy~(DOE) Office of Fossil Energy’s Carbon Storage Research Program via the Science-Informed Machine Learning to Accelerate Real-Time Decision Making for Carbon Storage~(SMART-CS) Initiative.

\clearpage

\bibliographystyle{unsrt}  

\clearpage
\appendix

Appendix arrangement:
\begin{itemize}
    \item \Cref{sup:license} provides an overview of the public resources available to facilitate reproducibility and outlines the licenses associated with the $\mathbf{\mathbb{E}^{FWI}}$ data and released code.
    \item \Cref{sup:forward} explains the particular method we utilized for elastic seismic forward modeling, which enabled us to generate the seismic data.
    \item \Cref{sup:naming} provides detailed information on the format and naming conventions used for $\mathbf{\mathbb{E}^{FWI}}$.
    \item \Cref{sup:network} provides a comprehensive description of the intricate architecture of ${\mathbb{E}}$lasticNet, ${\mathbb{E}}$lasticGAN, and ${\mathbb{E}}$lasticTransformer.
    \item \Cref{sup:training} outlines the specific training details employed for ${\mathbb{E}}$lasticNet, ${\mathbb{E}}$lasticGAN, and ${\mathbb{E}}$lasticTransformer.
    \item \Cref{sup:benchmark} showcases the benchmark results achieved by ${\mathbb{E}}$lasticGAN and ${\mathbb{E}}$lasticTransformer.
    \item \Cref{sec:ablation1} explores an ablation study focusing on the impacts of independently inverting $\mathbf{V_P}$ and $\mathbf{V_S}$, thereby sidelining their coupling effects, utilizing the $\mathbf{\mathbb{E}^{FWI}}$ datasets.
    \item \Cref{sec:ablation2} delves into an ablation study investigating the interdependencies and influence of variations in $\mathbf{V_P}$ and $\mathbf{V_S}$ structures on seismic data.
    \item \Cref{sec:computation} presents comprehensive computational details regarding elastic forward modeling.
    \item \Cref{sec:openfwi} presents a comparison between $\mathbf{\mathbb{E}^{FWI}}$ and \textsc{OpenFWI}.
\end{itemize}

\section{$\mathbf{\mathbb{E}^{FWI}}$ Public Resources and Licenses}
\label{sup:license}
To ensure the reliability of reproducing $\mathbf{\mathbb{E}^{FWI}}$ benchmarks, we have established several accessible resources. These resources are summarized in the following list. Additionally, our dedicated team is actively engaged in maintaining the platform and incorporating future advancements based on valuable feedback from the community.

\begin{itemize}
\item \textbf{Website: } \url{https://efwi-lanl.github.io}
\item \textbf{Dataset URL: } \url{ https://efwi-lanl.github.io/docs/data.html#vel}
\end{itemize}

Codes and pretrained model will be released upon approval by Los Alamos National Laboratory and U.S. Department of Energy.

\section{Seismic Data Generation}
\label{sup:forward}

In $\mathbf{\mathbb{E}^{FWI}}$, the seismic source and receiver geometries remain aligned with the \textsc{OpenFWI} dataset~\cite{deng2022openfwi}, except that the grid spacing is reduced to 5$m$ in order to preserve the stability of elastic wave propagation. Each velocity map is associated with a total of 5 seismic sources and 70 receivers, which are evenly distributed on the upper surface. This configuration ensures an abundance of source-receiver pairs for the purpose of seismic data generation.

Our Python forward modeling algorthm follows the Matlab code at~\url{https://csim.kaust.edu.sa/files/ErSE328.2013/LAB/Chapter.FD/lab.fdpsv/lab2.html}. The seismic data is simulated from the velocity maps using finite-difference solver~\cite{marfurt1984accuracy} with the elastic equations~\cite{levander1988fourth} with a 2nd-order accuracy in time and a 4th-order accuracy in space. A 350 grids absorbing boundary~\cite{engquist1977absorbing} is adopted to avoid the reflections from the model boundaries. This method provides a robust and computationally efficient mean of generating accurate seismic data that align with the $\mathrm{V_P}$ and $\mathrm{V_S}$ models. The point source function utilized in this study is a Ricker wavelet with a central frequency of 15 Hz. This particular wavelet is applied to the vertical component of particle displacement.

\section{$\mathbf{\mathbb{E}^{FWI}}$ Datasets: Illustration, Format, Naming}
\label{sup:naming}
\begin{figure*}[h!]
\centering
\includegraphics[width=1\columnwidth]{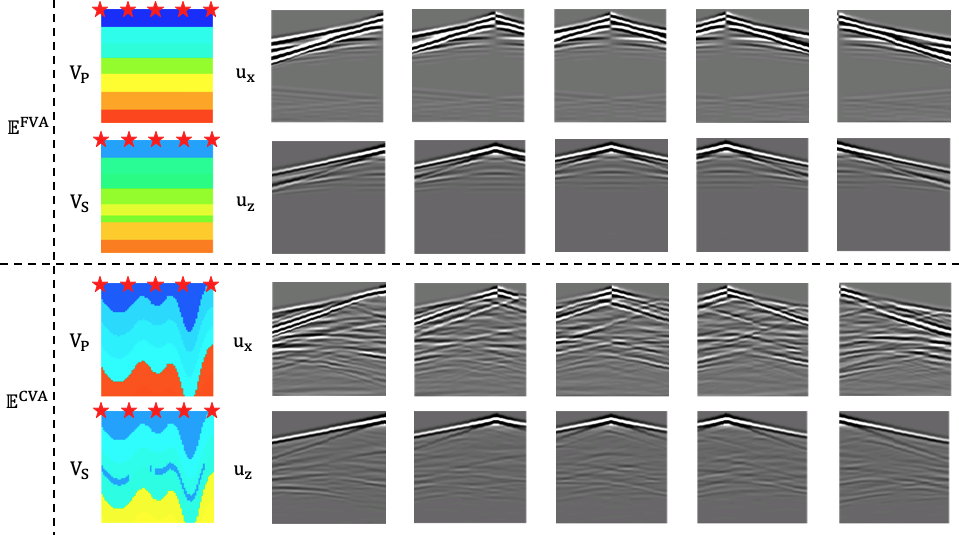}
\caption{\textbf{Examples of $\mathrm{V_P}$ and $\mathrm{V_S}$ maps, along with seismic data $u_x$ and $u_z$, in $\mathbf{\mathbb{E}^{FVA}}$ and $\mathbf{\mathbb{E}^{CVA}}$}. The star markers indicate the source locations.}
\label{fig:seismic_example1}
\end{figure*}

\begin{figure*}[h!]
\centering
\includegraphics[width=1\columnwidth]{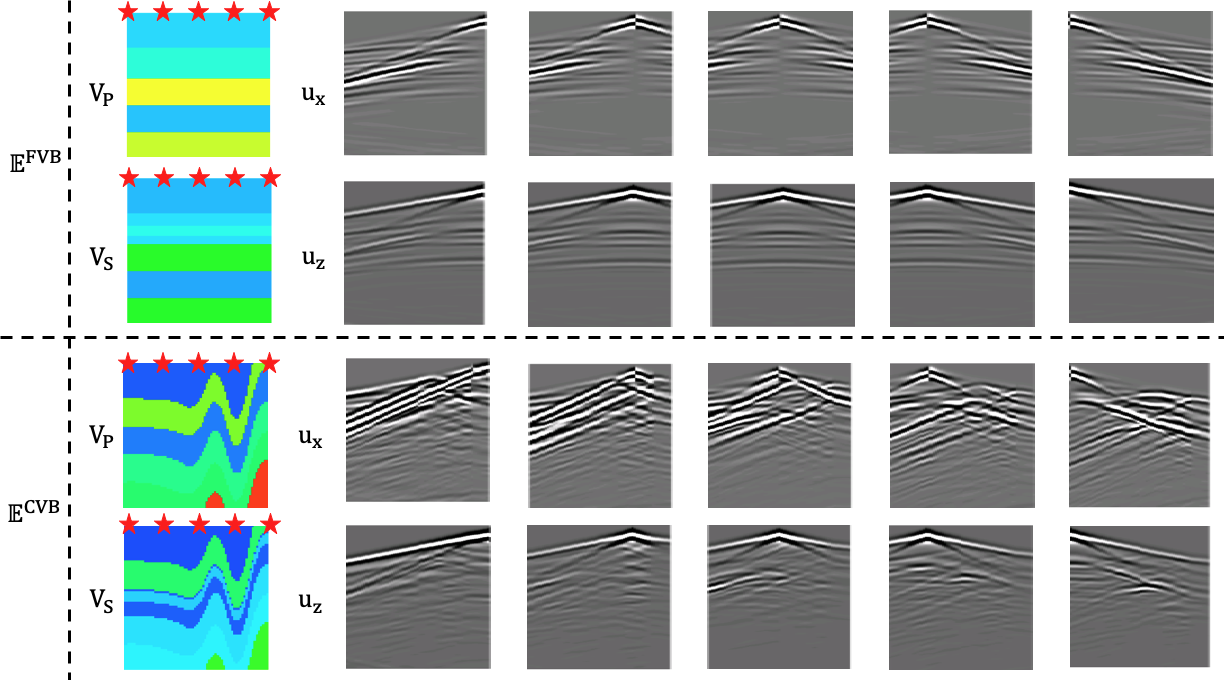}
\caption{\textbf{Examples of $\mathrm{V_P}$ and $\mathrm{V_S}$ maps, along with seismic data $u_x$ and $u_z$, in $\mathbf{\mathbb{E}^{FVB}}$ and $\mathbf{\mathbb{E}^{CVB}}$}. The star markers indicate the source locations.}
\label{fig:seismic_example3}
\end{figure*}

\begin{figure*}[h!]
\centering
\includegraphics[width=1\columnwidth]{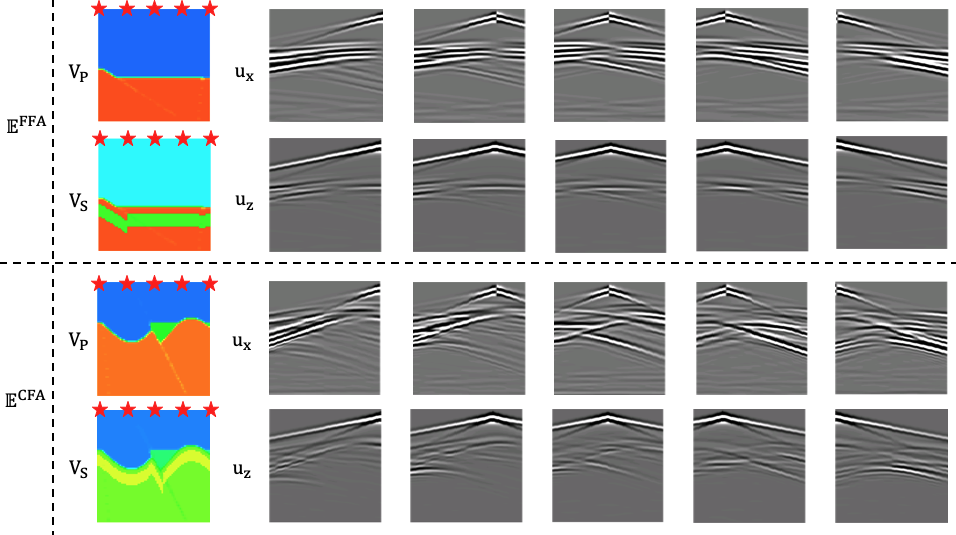}
\caption{\textbf{Examples of $\mathrm{V_P}$ and $\mathrm{V_S}$ maps, along with seismic data $u_x$ and $u_z$, in $\mathbf{\mathbb{E}^{FFA}}$ and $\mathbf{\mathbb{E}^{CFA}}$}. The star markers indicate the source locations.}
\label{fig:seismic_example2}
\end{figure*}

\begin{figure*}[h!]
\centering
\includegraphics[width=1\columnwidth]{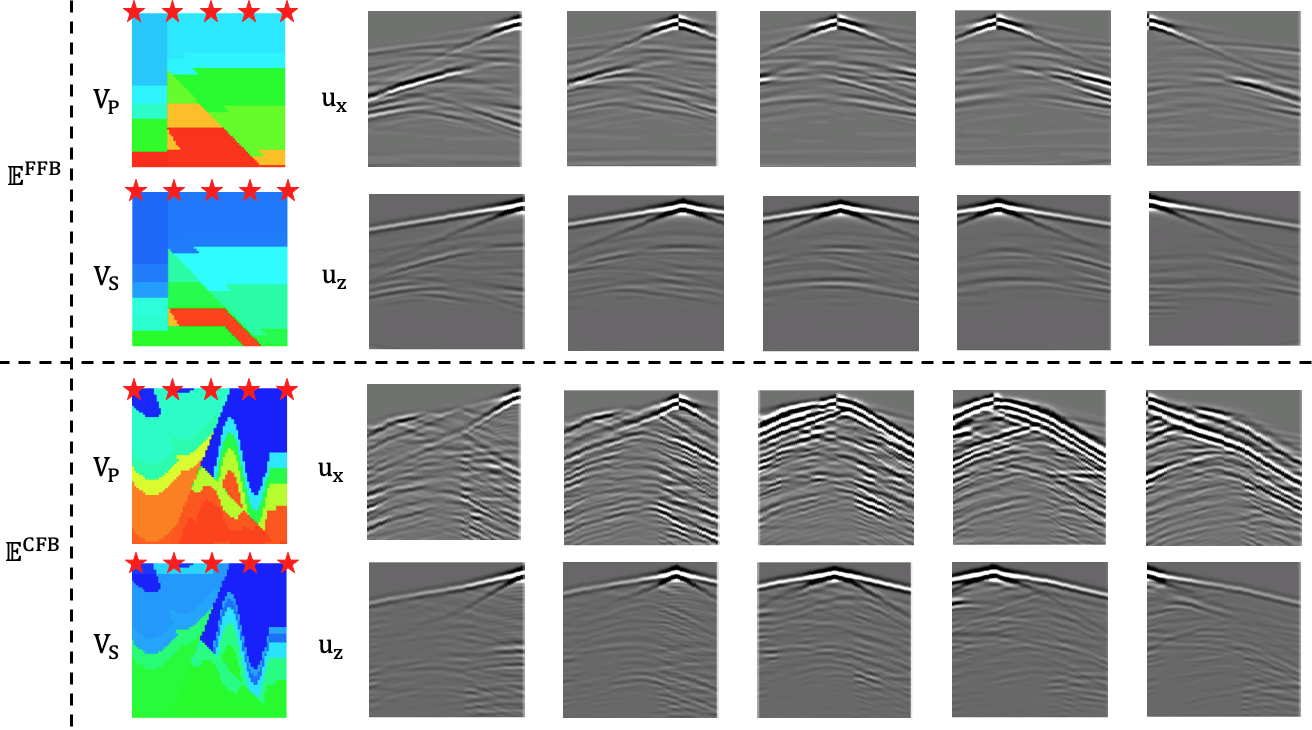}
\caption{\textbf{Examples of $\mathrm{V_P}$ and $\mathrm{V_S}$ maps, along with seismic data $u_x$ and $u_z$, in $\mathbf{\mathbb{E}^{FFB}}$ and $\mathbf{\mathbb{E}^{CFB}}$}. The star markers indicate the source locations.}
\label{fig:seismic_example4}
\end{figure*}

The $\mathbf{\mathbb{E}^{FWI}}$ datasets are organized into eight folders, namely {FVA}, {FVB}, {CVA}, {CVB}, {FFA}, {FFB}, {CFA}, and {CFB}. These folders contain the datasets $\mathbf{\mathbb{E}^{FVA}}$, $\mathbf{\mathbb{E}^{FVB}}$, $\mathbf{\mathbb{E}^{CVA}}$, $\mathbf{\mathbb{E}^{CVB}}$, $\mathbf{\mathbb{E}^{FFA}}$, $\mathbf{\mathbb{E}^{FFB}}$, $\mathbf{\mathbb{E}^{CFA}}$, and $\mathbf{\mathbb{E}^{CFB}}$, respectively. The examples of $\mathbf{V_P}$, $\mathbf{V_S}$ velocity maps, along with seismic data $u_x$ and $u_z$ in $\mathbf{\mathbb{E}^{FWI}}$ are shown in Figure~\ref{fig:seismic_example1} to~\ref{fig:seismic_example4}.

\textbf{Format:}
All samples in $\mathbf{\mathbb{E}^{FWI}}$ are stored in \texttt{.npy} format. The velocity maps, denoted as $\mathrm{V_P}$ and $\mathrm{V_S}$, as well as the Poisson's ratio $\mathrm{Pr}$, and the seismic data $u_x$ and $u_z$, are all stored separately in individual files for preservation. Each file contains a single NumPy array of 500 samples. The shapes of the arrays in velocity map files, Poisson's ratio, and seismic data files are (500, 1, 70, 70) and (500, 5, 1000, 70), respectively. 

\textbf{Naming:}
The naming of files can be described as \texttt{\{vp|vs|pr|data\_x|data\_z\}\_\{i\}.npy}, where \texttt{vp}, \texttt{vs}, \texttt{pr}, \texttt{data\_x} and \texttt{data\_z} specify whether a file stores $\mathrm{V_P}$, $\mathrm{V_S}$, $\mathrm{Pr}$, $u_x$ or $u_z$, \texttt{i} is the index of a file (start from 0) among the ones with the same \texttt{n}. Here are several examples:

\begin{itemize}
 \item \texttt{vp\_3.npy}  is the third file among all the files with $\mathrm{V_P}$ velocity maps. 
 \item \texttt{vs\_1.npy} is the first file among all the files with $\mathrm{V_S}$ velocity maps. 
\item \texttt{pr\_4.npy} is the fourth file among all the files with Poisson's ratio $\mathrm{Pr}$ maps. 
 \item \texttt{data\_x\_1.npy} is the file that contains the seismic data x component $u_x$ corresponding to the velocity maps in \texttt{vp\_1.npy} and  \texttt{vs\_1.npy}.
 \item \texttt{data\_z\_1.npy} is the file that contains the seismic data z component $u_z$ corresponding to the velocity maps in \texttt{vp\_1.npy} and  \texttt{vs\_1.npy}.
\end{itemize}

\section{$\mathbf{\mathbb{E}^{FWI}}$ Benchmarks: Network Architecture}
\label{sup:network}
\subsection{${\mathbb{E}}$lasticNet}
${\mathbb{E}}$lasticNet is an encoder-decoder structural CNN network built upon InversionNet~\cite{wu2019inversionnet}. The architecture consists of two encoders that take seismic data $u_x$ and $u_z$ as inputs, representing the horizontal and vertical components respectively. The encoder comprises a stack of 14 CNN layers. The first layer has a kernel size of $7\times1$, while the subsequent six layers have a kernel size of $3\times1$. To reduce the data dimension to the size of the velocity map, a stride of 2 is applied every other layer. Following this, six additional CNN layers with a kernel size of $3\times3$ are employed to extract spatial-temporal features from the compressed data. In these layers, the data is down-sampled every other layer using a stride of 2. Afterward, a CNN layer with an $8\times9$ kernel size is used to flatten the feature maps into a latent vector of size 512. The latent vectors from both encoders are concatenated and passed through two decoders to obtain P- and S-wave velocity maps, denoted as $\mathrm{V_P}$ and $\mathrm{V_S}$ respectively. In the decoder, the latent vector undergoes a deconvolutional layer to generate a $5\times5\times512$ tensor using a kernel size of 5. This is followed by a convolutional layer with the same number of input and output channels. This deconvolution-convolution process is repeated four times using a kernel size of 4 for the deconvolutional layers. As a result, a feature map of size $80\times80\times32$ is obtained. Finally, the feature map is center-cropped using a $70\times70$ window, and a $3\times3$ convolutional layer is applied to generate a single-channel velocity map. The overall architecture consists of 14 CNN layers in the encoder and 11 layers in the decoder. All the convolutional and deconvolutional layers are followed by batch normalization, and the activation function used is leakyReLU.


\subsection{${\mathbb{E}}$lasticGAN}

${\mathbb{E}}$lasticGAN is extended from the VelocityGAN architecture~\cite{zhang2020data} with an encoder-decoder CNN network as the generator, while the discriminator is a fully-CNN network. Note that the generator allows for other model architecture, though we adopt $\mathbb{E}$lasticNet  for the consistency on performance evaluation. The discriminator takes the generated velocity maps ($\mathrm{V_P}$, $\mathrm{V_S}$) as two inputs and classifies them into fake or true predictions. Each encoder of the discriminator has 9 convolution layers with LeakyReLU activation but not any normalization. The training process follows the common practice using the Wasserstein distance with gradient penalty, in addition to the pixel-wise $\ell_1$ or $\ell_2$ loss.

\subsection{${\mathbb{E}}$lasticTransformer} 

${\mathbb{E}}$lasticTransformer follows a similar seismic-encoder and velocity-decoder architecture design as the SimFWI described in \cite{feng2023simplifying}. 
It consists of two two-layer transformer encoders that take $u_x$ and $u_z$ as inputs and two two-layer transformer decoders to output $\mathrm{V_P}$ and $\mathrm{V_S}$ separately. The patch size of seismic data is $100\times10$, and the patch size of velocity maps is $10\times10$. The dimension of the encoder is $132$ with $12$ heads, and the dimension of the decoder is $516$ with $16$ heads. Similarly to SimFWI, we employ a linear layer to project the embedding of $u_x$ and $u_z$ into a $128$-dimensional space, separately. These projected embeddings are then concatenated and fed through two separate 2-piece Maxout layers to obtain the latent representations. Subsequently, two additional linear layers are utilized to map each latent representation to the appropriate dimensions of the respective decoders.
Unlike the linear upsampler utilized at the end of the velocity decoder in \cite{feng2023simplifying}, we stack four upsampling and $3\times3$ convolution blocks to construct the upsampler. Each block increases the size of the feature map by a factor of two and reduces the dimension by half. In the end there is another $3\times3$ convolution layer to generate the single-channel velocity map.

\section{$\mathbf{\mathbb{E}^{FWI}}$ Benchmarks: Training Configuration}
\label{sup:training}

This section presents the training configurations that have been implemented to ensure reproducibility. All experiments are conducted using NVIDIA Tesla V100 GPUs. We maintain consistent hyperparameters across all datasets for ${\mathbb{E}}$lasticNet, ${\mathbb{E}}$lasticGAN, and ${\mathbb{E}}$lasticTransformer. The AdamW optimizer~\cite{loshchilov2018decoupled} is employed with a weight decay of $1\times10^{-4}$ and momentum parameters $\beta_1=0.9$ and $\beta_2=0.999$ to update all models. During the training process, we apply min-max normalization to rescale the velocity maps and seismic data within the range of $[-1, 1]$. The velocity values for $\mathbf{V_P}$ maps range from $1500$ $m/s$ to $4500$ $m/s$, while the velocity values for $\mathbf{V_S}$ range from $612$ $m/s$ to $3000$ $m/s$. The learning rate is set to $1\times10^{-3}$ for ${\mathbb{E}}$lasticNet, ${\mathbb{E}}$lasticGAN and ${\mathbb{E}}$lasticTransformer. For ${\mathbb{E}}$lasticNet and ${\mathbb{E}}$lasticGAN, there are no weight decay and the batch size is set as 128. For ${\mathbb{E}}$lasticTransformer, the weight decay is 0.05 and the batch size is 256.

\section{$\mathbf{\mathbb{E}^{FWI}}$ Benchmarks: $\mathbb{E}$lasticGAN and $\mathbb{E}$lasticTransform} 
\label{sup:benchmark}
The benchmarks of $\mathbb{E}$lasticGAN and $\mathbb{E}$lasticTransform on $\mathbf{\mathbb{E}^{FWI}}$ are given in~\Cref{table:ElasticGAN_Benchmark} and~\Cref{table:ElasticTransformer_Benchmark}. $\mathbb{E}$lasticGAN demonstrates superior performance in predicting $\mathrm{V_P}$ and $\mathrm{V_S}$ compared to $\mathbb{E}$lasticNet, but for $\mathrm{Pr}$, the outcomes of $\mathbb{E}$lasticGAN are inferior to $\mathbb{E}$lasticNet. Among the three models, $\mathbb{E}$lasticTransform yields the best results for simple datasets such as $\mathbb{E}^{FVA}$, $\mathbb{E}^{FFA}$, and $\mathbb{E}^{CFA}$. However, as the complexity of velocity maps increases, $\mathbb{E}$lasticTransformer becomes the poorest performing model.
\begin{table}[!ht]
\footnotesize
\renewcommand{\arraystretch}{1.3}
\centering
\caption{\textbf{Quantitative results} of ElasticGAN on $\mathbf{\mathbb{E}^{FWI}}$ datasets. }
\vspace{0.5em}
\begin{adjustbox}{width=1\textwidth}
\begin{tabular}{c|cc|ccc|ccc|ccc} 
\thickhline
\multirow{3}{*}{Dataset}      & \multicolumn{2}{c|}{\multirow{3}{*}{Loss}} & \multicolumn{9}{c}{ElasticGAN}    \\ 
\cline{4-12}
    & \multicolumn{2}{c|}{}    & \multicolumn{3}{c|}{Vp}    & \multicolumn{3}{c|}{Vs}    & \multicolumn{3}{c}{Pr}    \\ 
\cline{4-12}
    & \multicolumn{2}{c|}{}    & \multicolumn{1}{c}{MAE$\downarrow$} & RMSE$\downarrow$ & SSIM$\uparrow$  & \multicolumn{1}{c}{MAE$\downarrow$} & RMSE$\downarrow$ & SSIM$\uparrow$  & \multicolumn{1}{c}{MAE$\downarrow$} & RMSE$\downarrow$ & SSIM$\uparrow$   \\ 
\hline
\multirow{2}{*}{$\mathbf{\mathbb{E}^{FVA}}$}
    & $\ell_1$ &
    & 0.0540    & 0.0882    & \textbf{0.9057} 
    & 0.0444    & 0.0809    & 0.8856 
    & 0.0571    & 0.1129    & \textbf{0.6814}    \\
    &    & $\ell_2$
    & 0.0506    & 0.0774    & 0.8736
    & 0.0403    & 0.0623    & \textbf{0.9016}
    & 0.0620    & 0.0960    & 0.5066    \\ 
\hline
\multirow{2}{*}{$\mathbf{\mathbb{E}^{FVB}}$}
    & $\ell_1$ &
    & 0.1087    & 0.2012    & \textbf{0.8064}
    & 0.0818    & 0.1520    & \textbf{0.8146}
    & 0.0757    & 0.1329    & \textbf{0.5806}    \\
    &    & $\ell_2$
    & 0.1177    & 0.1975    & 0.7887
    & 0.0815    & 0.1458    & 0.8080
    & 0.0817    & 0.1289    & 0.4764    \\ 
\hline
\multirow{2}{*}{$\mathbf{\mathbb{E}^{CVA}}$}
    & $\ell_1$ &    
    & 0.0983    & 0.1594    & \textbf{0.7661}
    & 0.0817    & 0.1331    & 0.7717
    & 0.0693    & 0.1258    & \textbf{0.5699}    \\
    &    & $\ell_2$
    & 0.1014    & 0.1524    & 0.7284
    & 0.0783    & 0.1204    & \textbf{0.7762}
    & 0.0847    & 0.1309    & 0.4398    \\ 
\hline
\multirow{2}{*}{$\mathbf{\mathbb{E}^{CVB}}$}
    & $\ell_1$ &
    & 0.1968    & 0.3293    & \textbf{0.6170}
    & 0.1469    & 0.2427    & \textbf{0.6470}
    & 0.0881    & 0.1505    & \textbf{0.4622}    \\
    &    & $\ell_2$
    & 0.2077    & 0.3218    & 0.6069
    & 0.1574    & 0.2386    & 0.6237
    & 0.1067    & 0.1563    & 0.3639    \\ 
\hline
\multirow{2}{*}{$\mathbf{\mathbb{E}^{FFA}}$}
    & $\ell_1$ &
    & 0.0846    & 0.1452    & 0.8464
    & 0.0725    & 0.1204    & 0.8447
    & 0.0812    & 0.1394    & 0.5954    \\
    &    & $\ell_2$
    & 0.0598    & 0.1015    & \textbf{0.8864}
    & 0.0494    & 0.0833    & \textbf{0.8883} 
    & 0.0592    & 0.1017    & \textbf{0.6206}    \\ 
\hline
\multirow{2}{*}{$\mathbf{\mathbb{E}^{FFB}}$}
    & $\ell_1$ &
    & 0.1177    & 0.1781    & \textbf{0.6992}
    & 0.0883    & 0.1373    & \textbf{0.7411}
    & 0.0543    & 0.1016    & \textbf{0.6347}    \\
    &    & $\ell_2$
    & 0.1268    & 0.1798    & 0.6387
    & 0.0921    & 0.1338    & 0.7386
    & 0.0690    & 0.1102    & 0.4747    \\ 
\hline
\multirow{2}{*}{$\mathbf{\mathbb{E}^{CFA}}$}
    & $\ell_1$ &
    & 0.0813    & 0.1471    & 0.8291
    & 0.0683    & 0.1234    & 0.8164
    & 0.0567    & 0.1179    & \textbf{0.6601}    \\
    &    & $\ell_2$
    & 0.0736    & 0.1176    & \textbf{0.8311}
    & 0.0602    & 0.0989    & \textbf{0.8479}
    & 0.0716    & 0.1148    & 0.5066    \\ 
\hline
\multirow{2}{*}{$\mathbf{\mathbb{E}^{CFB}}$}
    & $\ell_1$ &
    & 0.1639    & 0.2366    & \textbf{0.6096}
    & 0.1199    & 0.1741    & \textbf{0.6549}
    & 0.0621    & 0.1055    & 0.5780    \\
    &    & $\ell_2$
    & 0.1639    & 0.2326    & 0.6064
    & 0.1208    & 0.1726    & \textbf{0.6549}
    & 0.0593    & 0.1009    & \textbf{0.6076}    \\
\thickhline
\end{tabular}
\end{adjustbox}
\vspace{0.5em}
\label{table:ElasticGAN_Benchmark}
\vspace{-1em}
\end{table}

\begin{table}[!ht]
\footnotesize
\renewcommand{\arraystretch}{1.3}
\centering
\caption{\textbf{Quantitative results} of $\mathbb{E}$lasticTransformer on $\mathbf{\mathbb{E}^{FWI}}$ datasets. }
\vspace{0.5em}
\begin{adjustbox}{width=1\textwidth}
\begin{tabular}{c|cc|ccc|ccc|ccc} 
\thickhline
\multirow{3}{*}{Dataset}      & \multicolumn{2}{c|}{\multirow{3}{*}{Loss}} & \multicolumn{9}{c}{$\mathbb{E}$lasticTransformer}\\ 
\cline{4-12}
& \multicolumn{2}{c|}{}    & \multicolumn{3}{c|}{Vp}    & \multicolumn{3}{c|}{Vs}    & \multicolumn{3}{c}{Pr}    \\ 
\cline{4-12}
& \multicolumn{2}{c|}{}    & \multicolumn{1}{c}{MAE$\downarrow$} & RMSE$\downarrow$ & SSIM$\uparrow$  & \multicolumn{1}{c}{MAE$\downarrow$} & RMSE$\downarrow$ & SSIM$\uparrow$  & \multicolumn{1}{c}{MAE$\downarrow$} & RMSE$\downarrow$ & SSIM$\uparrow$   \\ 
\hline
\multirow{2}{*}{$\mathbf{\mathbb{E}^{FVA}}$}
    & $\ell_1$ &
    & 0.0326    & 0.0676    & 0.9359
    & 0.0232    & 0.0514    & 0.9386
    & 0.0351    & 0.0772    & \textbf{0.7891} \\
    &    & $\ell_2$
    & 0.0337    & 0.0670    & \textbf{0.9389}
    & 0.0240    & 0.0511    & \textbf{0.9413}
    & 0.0367    & 0.0761    & 0.7810    \\ 
\hline
\multirow{2}{*}{$\mathbf{\mathbb{E}^{FVB}}$}
    & $\ell_1$ &
    & 0.0830    & 0.1794    & \textbf{0.8510}
    & 0.0595    & 0.1269    & \textbf{0.8609}
    & 0.0692    & 0.1303    & \textbf{0.6464} \\
    &    & $\ell_2$ 
    & 0.0871    & 0.1810    & 0.8466
    & 0.0641    & 0.1305    & 0.8531
    & 0.0723    & 0.1319    & 0.6166    \\
\hline
\multirow{2}{*}{$\mathbf{\mathbb{E}^{CVA}}$}
    & $\ell_1$ &
    & 0.0826    & 0.1448    & 0.8000
    & 0.0648    & 0.1109    & 0.7967
    & 0.0959    & 0.1450    & 0.4650    \\
    &    & $\ell_2$
    & 0.0853    & 0.1398    & \textbf{0.8068}
    & 0.0659    & 0.1070    & \textbf{0.8103}
    & 0.1034    & 0.1467    & \textbf{0.4633} \\ 
\hline
\multirow{2}{*}{$\mathbf{\mathbb{E}^{CVB}}$}
    & $\ell_1$ &
    & 0.1772    & 0.3129    & 0.6548
    & 0.1294    & 0.2249    & 0.6777
    & 0.1225    & 0.1983    & \textbf{0.3934} \\
    &    & $\ell_2$
    & 0.1838    & 0.2933    & \textbf{0.6670}
    & 0.1354    & 0.2144    & \textbf{0.6898} 
    & 0.1363    & 0.1951    & 0.3588 \\
\hline
\multirow{2}{*}{$\mathbf{\mathbb{E}^{FFA}}$}
    & $\ell_1$ &
    & 0.1190    & 0.1765    & 0.8827
    & 0.0779    & 0.1178    & 0.8513
    & 0.1544    & 0.1898    & 0.6129 \\
    &    & $\ell_2$
    & 0.1153    & 0.1634    & \textbf{0.8868}
    & 0.0671    & 0.1023    & \textbf{0.8691} 
    & 0.1418    & 0.1780    & \textbf{0.6322}    \\
\hline
\multirow{2}{*}{$\mathbf{\mathbb{E}^{FFB}}$} 
    & $\ell_1$ & 
    & 0.1120    & 0.1760    & 0.7048
    & 0.0802    & 0.1277    & 0.7492
    & 0.0960    & 0.1478    & 0.4316    \\
    &    & $\ell_2$ 
    & 0.1161    & 0.1716    & \textbf{0.7272} 
    & 0.0819    & 0.1247    & \textbf{0.7644}
    & 0.0886    & 0.1330    & \textbf{0.4950} \\
\hline
\multirow{2}{*}{$\mathbf{\mathbb{E}^{CFA}}$}
    & $\ell_1$ &
    & 0.0372    & 0.0924    & 0.9100
    & 0.0365    & 0.0878    & 0.8768
    & 0.0422    & 0.1048    & \textbf{0.7003} \\
    &    & $\ell_2$ 
    & 0.0601    & 0.1100    & \textbf{0.8961}
    & 0.0498    & 0.0909    & \textbf{0.8710} 
    & 0.0621    & 0.1127    & 0.6258    \\
\hline
\multirow{2}{*}{$\mathbf{\mathbb{E}^{CFB}}$}
    & $\ell_1$ &
    & 0.1863    & 0.2727    & 0.5560
    & 0.1388    & 0.2045    & 0.6002
    & 0.1433    & 0.2101    & 0.3203    \\
    &    & $\ell_2$ 
    & 0.1787    & 0.2532    & \textbf{0.5750}
    & 0.1302    & 0.1863    & \textbf{0.6283} 
    & 0.1284    & 0.1842    & \textbf{0.3578} \\
\thickhline
\end{tabular}
\end{adjustbox}
\vspace{0.5em}
\label{table:ElasticTransformer_Benchmark}
\vspace{-1em}
\end{table}

\section{Independent vs. Joint Inversion: Impact on $\mathbf{\mathrm{Pr}}$ Maps}
\label{sec:ablation1}


This experiment seeks to explore the implications of independent versus joint inversion of $\mathrm{V_P}$ and $\mathrm{V_S}$ on the precision of predicted $\mathrm{Pr}$ maps, thereby underscoring the importance of considering the relationship between $\mathrm{V_P}$ and $\mathrm{V_S}$, as well as the coupling of P and S waves.

\begin{figure*}[h!]
\centering
\includegraphics[width=1\columnwidth]{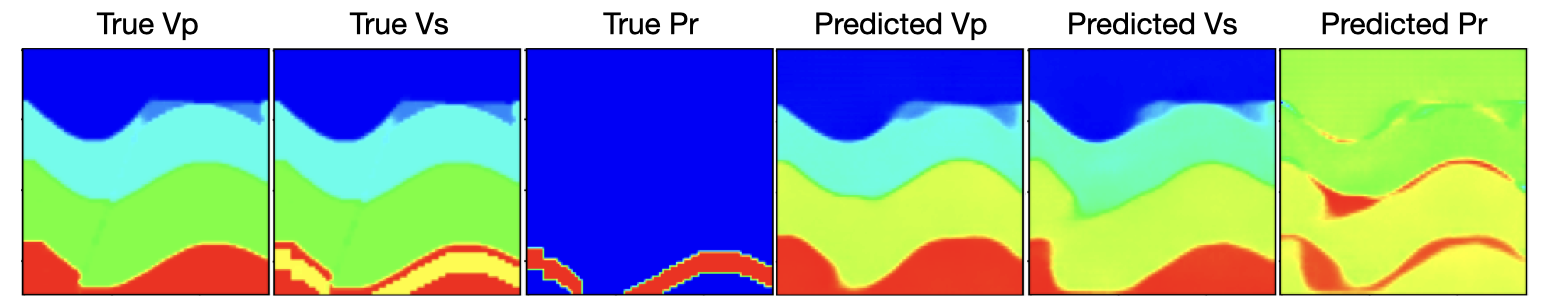}
\caption{\textbf{Examples of independent inversion results of $\mathbf{\mathbb{E}^{CFA}}$ dataset:} from left to right: ground truth $\mathrm{V_P}$, ground truth $\mathrm{V_S}$, ground truth $\mathrm{P_r}$, predicted $\mathrm{V_P}$, predicted $\mathrm{V_S}$, and predicted $\mathrm{P_r}$.}
\label{fig:independent_vel}
\end{figure*}

The procedure involves the individual training of two separate InversionNets utilizing the $\mathbf{\mathbb{E}^{FWI}}$ dataset. One InversionNet is tasked with the prediction of $\mathrm{V_P}$ maps, while the other focuses on the prediction of $\mathrm{V_S}$ maps. Subsequently, the independently predicted $\mathrm{V_P}$ and $\mathrm{V_S}$ outputs are used to compute the Poisson's ratio maps. The derived maps are then statistically compared to the ground truth $\mathrm{Pr}$ maps. Further details of these metrics are provided in Table 5 in the main article.

The outcome reveals a significantly higher MAE and MSE, coupled with lower SSIM values for $\mathrm{Pr}$ maps reconstructed from independent $\mathrm{V_P}$/$\mathrm{V_S}$ predictions when juxtaposed with those reconstructed from joint inversion (Table 3 in the main article). Specifically, the SSIM for the $\mathrm{P_r}$ map reconstructed from independent inversion in the "$\mathbf{\mathbb{E}^{FVA}}$" set is $2\%$ less than the ${\mathbb{E}}$lasticNet joint inversion result in the $\ell_1$ case and $6\%$ in the $\ell_2$ case. Even in comparison to the InversionNet results in Table 5 in the main article, the Pr performance decrement is still significant. Additionally, both the MAE and MSE of the independent inversion exceed those of the joint inversion result by $60\%$ and by $72\%$ of the independent same structural inversion result. An example visualized comparison is shown in~\Cref{fig:independent_vel}. Of critical importance is the observation that key targets of the inversion, the reservoir thin layers, are misplaced at incorrect depths and exhibit distorted spatial shapes in the independent inversion. This misrepresentation could lead to erroneous reservoir exploration, with potentially severe economic ramifications.

Consequently, this experiment reinforces the necessity of incorporating the $\mathrm{V_P}$-$\mathrm{V_S}$ relationship and the P-S wave coupling in the imaging and targeting of reservoirs. Ignoring the elastic wave coupling and focusing solely on single-parameter inversion is deemed unviable.

\section{Investigating P- and S-waves Coupling via Machine Learning}
\label{sec:ablation2}

\begin{figure*}[h!]
\centering
\includegraphics[width=0.85\columnwidth]{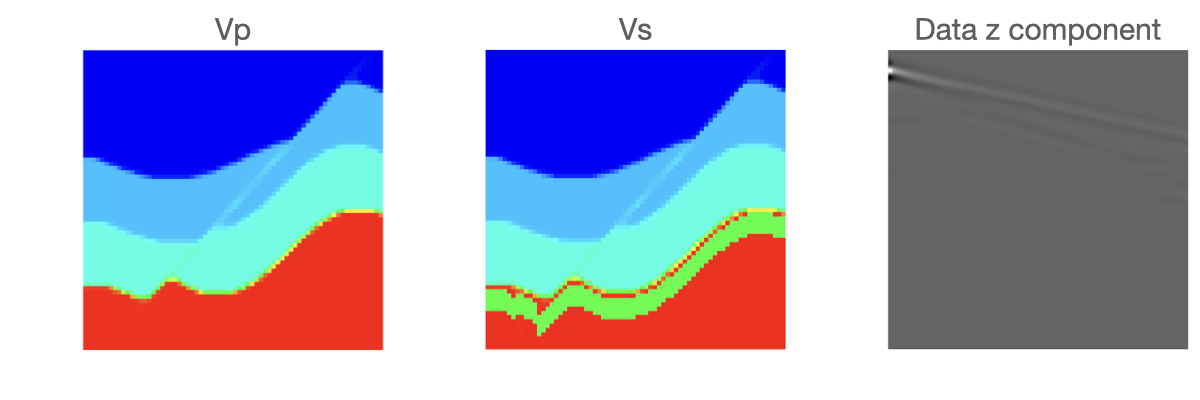}
\caption{\textbf{Examples of independent inversion of $\mathbf{\mathbb{E}^{CFA}}$ training set:} from left to right: ground truth $\mathrm{V_P}$, ground truth $\mathrm{V_S}$, ground truth seismic data z-component $\mathrm{u_Z}$. All used for independent $\mathrm{V_P}$, $\mathrm{V_S}$ trainings.}
\label{fig:ablation_8_true}
\end{figure*}

The primary objective of this experiment is to elucidate the coupling effects of P and S waves in seismic data. Specifically, we evaluate the feasibility of a $\mathrm{V_P}$-only data-driven FWI that overlooks $\mathrm{V_S}$ structural alterations. The experimental design is detailed below:

\textbf{Training:} We utilize \textsc{OpenFWI}'s InversionNet, which is trained using the Z-component seismic data sourced from $\mathbf{\mathbb{E}^{FWI}}$. This setup is identical to the benchmarks set by the \textsc{OpenFWI}'s InversionNet, with the exception that our input seismic data incorporates elastic effects. The output is confined to the $\mathrm{V_P}$ maps. We utilize $48,000$ and $24,000$ training samples for the "$\mathbb{E}^{Fault}$" and "$\mathbb{E}^{Vel}$" families, respectively, and reserve the remaining samples for testing. Training samples of the "$\mathbb{E}^{CFA}$" set are shown in~\Cref{fig:ablation_8_true}.

\begin{figure*}[h!]
\centering
\includegraphics[width=1\columnwidth]{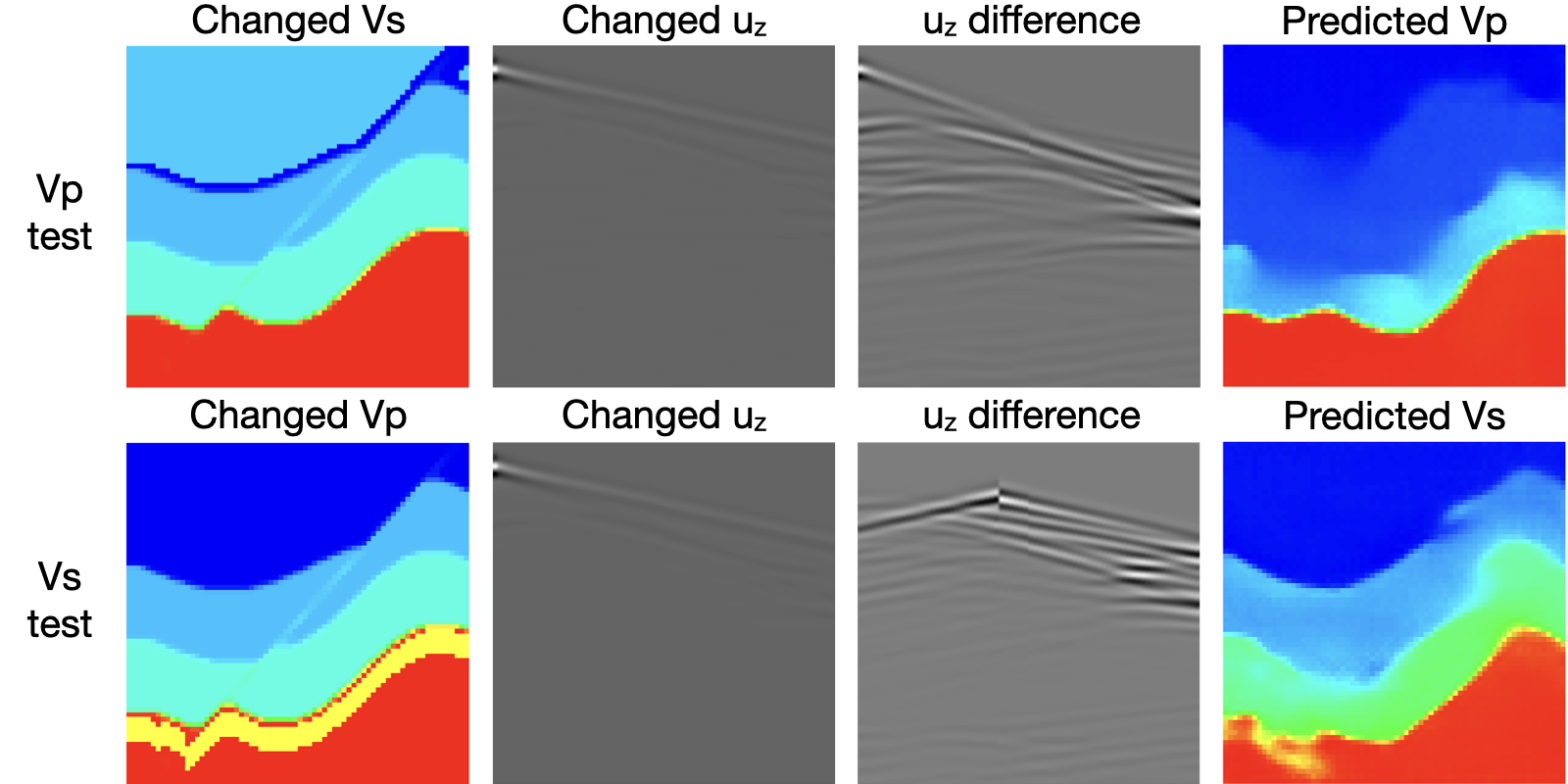}
\caption{\textbf{Examples of independent inversion results of changed structural $\mathbf{\mathbb{E}^{CFA}}$ dataset:} from left to right: changed structural velocities $\mathrm{V_S}$($\mathrm{V_P}$) for independent $\mathrm{V_P}$($\mathrm{V_S}$) test, corresponding data z-component $\mathrm{u_Z}$ with changed $\mathrm{V_S}$($\mathrm{V_P}$), corresponding data difference, predicted $\mathrm{V_P}$($\mathrm{V_S}$) with changed $\mathrm{u_Z}$ as input.}
\label{fig:ablation_8_pred}
\end{figure*}

\textbf{Testing:} The testing phase is divided into two steps: 1) We use the aforementioned reserved test sets as a benchmark to evaluate the performance of InversionNet. 2) We then generate a new elastic seismic dataset by eradicating the thin layer reservoir structure in the Poisson maps, leading to paired $\mathrm{V_P}$/$\mathrm{V_S}$ maps with identical structures, shown in~\Cref{fig:ablation_8_pred}. 
%
This new dataset is identical to the original $\mathbf{\mathbb{E}^{FWI}}$ dataset, save for the minor alterations in $\mathrm{V_S}$ structure and the corresponding changes in seismic data. Consequently, the performance differences between the two testing sets should solely reflect the effects of differing $\mathrm{V_S}$ structures.

Our observations show a discernible performance decline when testing datasets with differing $\mathrm{V_S}$ structures. For instance, in the "$\mathbf{\mathbb{E}^{FVA}}$" set, the new dataset averagely increase by $343\%$ in MAE, and $296\%$ in RMSE compared to the baseline, while SSIM drops by $7\%$ in both $\ell_1$ and $\ell_2$ cases. A detailed account of these statistical metrics is provided in the main article Table 5 and 6.

Conversely, we performed a reciprocal experiment: training on $\mathrm{V_S}$-only and testing using the $\mathbf{\mathbb{E}^{FWI}}$ dataset, followed by further testing with altered $\mathrm{V_P}$ structures in the thin reservoir layer.
The observed outcomes mirror those of the first experiment. With a $\mathrm{V_S}$-only InversionNet, when $\mathrm{V_P}$ structures are altered slightly in the testing sets, the network's performance notably diminishes. For instance, in the complex case like the "$\mathbf{\mathbb{E}^{CFB}}$" set, the MAE and RMSE for the new dataset increase by $120\%$ and $89\%$ compared to the baseline in the $\ell_1$ case and increase by $115\%$ and $88\%$ in the $\ell_2$ cases, respectively, while SSIM decreases by $7\%$ in both $\ell_1$ and $\ell_2$ cases, respectively. The decrement is even large when compared to the $\mathbb{E}$lasticGAN (\Cref{table:ElasticGAN_Benchmark}) and the $\mathbb{E}$lasticTransform (\Cref{table:ElasticTransformer_Benchmark}) benchmarks. It worth mentioning that the diminished Pr prediction performance is based on the same level of Vp and Vs predictions for acoustic and elastic cases. Furthermore, the reservoir layer is not distinctly identifiable in the Pr maps, which could result in significant errors in reservoir estimation, potentially causing substantial economic loss. Detailed statistical metrics are listed in the main article Table 5 and 6.

In summary, single-parameter data-driven inversion, which neglects the coupling of P and S waves, results in substantial degradation of inversion performance. Consequently, the simultaneous prediction of $\mathrm{V_P}$ and $\mathrm{V_S}$ by considering the coupling of P and S waves proves to be indispensable.

\section{Computational Cost in Elastic Forward Modeling}
\label{sec:computation}

The adoption of the elastic approximation reintroduces the concern of computational costs. In the context of data-driven elastic FWI, a significant portion of the computational expenses arises from the construction of the training set. 
In contrast to elastic forward modeling, which generates particle displacement components $u_x$ and $u_z$ from velocity maps $\mathrm{V_P}$ and $\mathrm{V_S}$, acoustic forward modeling focuses solely on generating stress $p$ from the P-wave velocity map $\mathrm{V_P}$. The acoustic forward modeling process is governed by the acoustic wave equation, which can be expressed as follows:
\begin{equation}
\nabla^2p-\frac{1}{{V_{P}}^2}\frac{\partial ^2p}{\partial t ^2}=s,
\label{Acoustic}
\end{equation}
where $\nabla^2=\frac{\partial ^2}{\partial x ^2}+\frac{\partial ^2}{\partial z ^2}$, $\mathrm{V_P}$ is P-wave velocity map, $p$ is pressure field and $s$ is source term. 

When comparing elastic approximation to acoustic approximation, the generation of seismic data imposes substantially higher memory and computational burdens across multiple factors~\cite{hobro2014method}. 

\begin{itemize}
\item \textbf{Velocity Maps:} In an acoustic medium, only the P-wave velocity is sufficient to characterize the properties of the medium at a specific location. However, in an elastic medium, two parameters (the P-wave and S-wave velocities), are needed for an accurate description.
\item \textbf{Seismic data:} Seismic data in the domain of an elastic medium consists of the stress tensor encompassing horizontal and vertical components. Conversely, seismic data within an acoustic scenario primarily encompasses pressure, denoting a scalar quantity. Thus The memory required to store the elastic wavefield is at least twice of the acoustic wavefield.
\item \textbf{Wave Equation:}  The computational burden associated with solving the equation of motion and constitutive equations is notably reduced in acoustic modeling compared to elastic modeling. Specifically, the computational cost of elastic modeling is found to be three to six times higher than that of acoustic modeling when both are implemented on an identical grid.
\item \textbf{Stability:} In order to mitigate numerical dispersion, it is necessary for the grid size to correspond to the minimum velocity within the model, with the minimum Vs for elastic cases, and the minimum Vp for acoustic cases. Consequently, elastic modeling requires a finer grid spacing compared to acoustic modeling. In the context of 2D simulations, the relationship between the number of times the wave equation needs to be solved for acoustic simulations ($N_{acoustic}$) and elastic simulations ($N_{elastic}$) is expressed as follows:

\begin{equation}
\frac{N_{elastic}}{N_{acoustic}}=\left(\frac{V^{min}_{P}}{V^{min}_{S}}\right)^3
\end{equation}
The ratio between $\mathbf{V_P}$ and $\mathbf{V_S}$ usually ranges from 1.4 to 2.1~\cite{brantut2019influence,wang2012high,alam2017near}, which make $\frac{N_{elastic}}{N_{acoustic}}$ ranges from 2.7 to.9.3.

\end{itemize}
Elastic wave propagation simulations generally necessitate at least twice the memory compared to acoustic simulations. The computational demands can range from approximately 4.2 to 55.8 times higher. Consequently, the computation and memory requirements for generating $\mathbf{\mathbb{E}^{FWI}}$ are significantly greater than those of \textsc{OpenFWI}.

\section{$\mathbf{\mathbb{E}^{FWI}}$ vs \textsc{OpenFWI}}
\label{sec:openfwi}

While $\mathbf{\mathbb{E}^{FWI}}$ is constructed upon the foundations of \textsc{OpenFWI}, there remain notable distinctions between these two datasets. Beyond the computational expenses associated with the wave equation discussed in \Cref{sec:computation},~\Cref{table:openfwi} enumerates additional contrasts. Despite the fact that the data size of $\mathbf{\mathbb{E}^{FWI}}$ is smaller than that of \textsc{OpenFWI}, the inherent complexity of the elastic wave equation makes solving for both $\mathbf{V_P}$ and $\mathbf{V_S}$ from particle displacements $u_x$ and $u_z$ significantly more challenging than solving solely for $\mathbf{V_P}$ from pressure $p$.

\begin{table}
\centering
\caption{Dataset comparison between $\mathbf{\mathbb{E}^{FWI}}$ and \textsc{OpenFWI}}
\label{table:openfwi}
\begin{tabular}{c|c|c} 
\thickhline
\textbf{Dataset} & $\mathbf{\mathbb{E}^{FWI}}$                      & \textbf{\textsc{OpenFWI} }              \\ 
\thickhline

Dataset Families & $\mathbf{\mathbb{E}^{Vel}}$, $\mathbf{\mathbb{E}^{Fault}}$                          & Vel, Fault, Style, Kimberlina  \\
\hline
Wave Equation    & Elastic                             & Acoustic                       \\
\hline

Total Size       & 0.69 TB                             & 1.83 TB                       \\
\hline

Total Sample     & 168 K                               & 256 K                          \\
\hline

Input            & Particle displacement $u_x$ and $u_z$~                      & ~Pressure $p$                     \\
\hline

Output           & $\mathrm{V_P}$ and $\mathrm{V_S}$                       & $\mathrm{V_P}$                          \\
\hline

Target           & Obtain decoupled $\mathrm{V_P}$ and $\mathrm{V_S}$ maps & Obtain accurate $\mathrm{V_P}$ maps      \\
\thickhline
\end{tabular}
\end{table}


\begin{thebibliography}{10}

\bibitem{Virieux-2009-Overview}
J.~Virieux and S.~Operto.
\newblock An overview of full-waveform inversion in exploration geophysics.
\newblock {\em Geophysics}, 74(6):WCC1--WCC26, 2009.

\bibitem{virieux2017introduction}
Jean Virieux, Amir Asnaashari, Romain Brossier, Ludovic M{\'e}tivier,
  Alessandra Ribodetti, and Wei Zhou.
\newblock An introduction to full waveform inversion.
\newblock In {\em Encyclopedia of exploration geophysics}, pages R1--1. Society
  of Exploration Geophysicists, 2017.

\bibitem{fichtner2011resolution}
Andreas Fichtner and Jeannot Trampert.
\newblock Resolution analysis in full waveform inversion.
\newblock {\em Geophysical Journal International}, 187(3):1604--1624, 2011.

\bibitem{fichtner2013multiscale}
Andreas Fichtner, Jeannot Trampert, Paul Cupillard, Erdinc Saygin, Tuncay
  Taymaz, Yann Capdeville, and Antonio Villasenor.
\newblock Multiscale full waveform inversion.
\newblock {\em Geophysical Journal International}, 194(1):534--556, 2013.

\bibitem{vigh2011full}
Denes Vigh, Jerry Kapoor, and Hongyan Li.
\newblock Full-waveform inversion application in different geological settings.
\newblock In {\em 2011 SEG Annual Meeting}. OnePetro, 2011.

\bibitem{pratt1999seismic}
R~Gerhard Pratt.
\newblock Seismic waveform inversion in the frequency domain, part 1: Theory
  and verification in a physical scale model.
\newblock {\em Geophysics}, 64(3):888--901, 1999.

\bibitem{barnes2009domain}
Christophe Barnes and Marwan Charara.
\newblock The domain of applicability of acoustic full-waveform inversion for
  marine seismic data.
\newblock {\em Geophysics}, 74(6):WCC91--WCC103, 2009.

\bibitem{hobro2014method}
James~WD Hobro, Chris~H Chapman, and Johan~OA Robertsson.
\newblock A method for correcting acoustic finite-difference amplitudes for
  elastic effects.
\newblock {\em Geophysics}, 79(4):T243--T255, 2014.

\bibitem{raknes2017challenges}
Espen~Birger Raknes and B{\o}rge Arntsen.
\newblock Challenges and solutions for performing 3d time-domain elastic
  full-waveform inversion.
\newblock {\em The leading edge}, 36(1):88--93, 2017.

\bibitem{agudo2018acoustic}
{\`O}scar~Calder{\'o}n Agudo, Nuno~Vieira da~Silva, Michael Warner, and Joanna
  Morgan.
\newblock Acoustic full-waveform inversion in an elastic world.
\newblock {\em Geophysics}, 83(3):R257--R271, 2018.

\bibitem{fu2018multiscale}
Lei Fu, Bowen Guo, and Gerard~T Schuster.
\newblock Multiscale phase inversion of seismic data.
\newblock {\em Geophysics}, 83(2):R159--R171, 2018.

\bibitem{fang2020effects}
Jinwei Fang, Hui Zhou, Qingchen Zhang, Hanming Chen, Pengyuan Sun, Jianlei
  Zhang, and Liang Zhang.
\newblock The effects of elastic data on acoustic and elastic full waveform
  inversion.
\newblock {\em Journal of Applied Geophysics}, 172:103876, 2020.

\bibitem{feng2021mpi}
Shihang Feng, Lei Fu, Zongcai Feng, and Gerard~T Schuster.
\newblock Multiscale phase inversion for vertical transverse isotropic media.
\newblock {\em Geophysical Prospecting}, 69(8-9):1634--1649, 2021.

\bibitem{sears2008elastic}
Timothy~J Sears, SC~Singh, and PJ~Barton.
\newblock Elastic full waveform inversion of multi-component obc seismic data.
\newblock {\em Geophysical Prospecting}, 56(6):843--862, 2008.

\bibitem{liu2019innovative}
Xiaoqiang Liu, Zhanqing Qu, Tiankui Guo, Qizhong Tian, Wei Lv, Zhishuang Xie,
  and Chunbo Chu.
\newblock An innovative technology of directional propagation of hydraulic
  fracture guided by radial holes in fossil hydrogen energy development.
\newblock {\em International Journal of Hydrogen Energy}, 44(11):5286--5302,
  2019.

\bibitem{lees2000poisson}
Jonathan~M Lees and Huatao Wu.
\newblock Poisson's ratio and porosity at coso geothermal area, california.
\newblock {\em Journal of volcanology and geothermal research},
  95(1-4):157--173, 2000.

\bibitem{feng2020lithofacies}
Runhai Feng, Niels Balling, and Dario Grana.
\newblock Lithofacies classification of a geothermal reservoir in denmark and
  its facies-dependent porosity estimation from seismic inversion.
\newblock {\em Geothermics}, 87:101854, 2020.

\bibitem{li2021elastic}
Hui Li, Jing Lin, Baohai Wu, Jinghuai Gao, and Naihao Liu.
\newblock Elastic properties estimation from prestack seismic data using ggcnns
  and application on tight sandstone reservoir characterization.
\newblock {\em IEEE Transactions on Geoscience and Remote Sensing}, 60:1--21,
  2021.

\bibitem{Eugenia-vpvs}
Eugenia Rojas, Thomas~L. Davis, Michael Batzle, Manika Prasad, and Reinaldo~J.
  Michelena.
\newblock $\mathrm{V_p}$-$\mathrm{V_s}$ ratio sensitivity to pressure, fluid,
  and lithology changes in tight gas sandstones.
\newblock In {\em SEG Technical Program Expanded Abstracts 2005}, pages
  1401--1404, 2005.

\bibitem{Ding-vpvs}
Pinbo Ding, Ding Wang, Guidong Di, and Xiangyang Li.
\newblock Investigation of the effects of fracture orientation and saturation
  on the vp/vs ratio and their implications.
\newblock {\em Rock Mechanics and Rock Engineering}, 52:3293--3304, 2019.

\bibitem{Sun-vpvs}
J.~Sun, X.~Wei, and X.~Chen.
\newblock {Fluid identification in tight sandstone reservoirs based on a new
  rock physics model}.
\newblock {\em Journal of Geophysics and Engineering}, 13(4):526--535, 07 2016.

\bibitem{mulder2008exploring}
WA~Mulder and R-E Plessix.
\newblock Exploring some issues in acoustic full waveform inversion.
\newblock {\em Geophysical Prospecting}, 56(6):827--841, 2008.

\bibitem{operto2013guided}
St{\'e}phane Operto, Yaser Gholami, Vincent Prieux, Alessandra Ribodetti,
  R~Brossier, L~Metivier, and Jean Virieux.
\newblock A guided tour of multiparameter full-waveform inversion with
  multicomponent data: From theory to practice.
\newblock {\em The leading edge}, 32(9):1040--1054, 2013.

\bibitem{zhang2020high}
Zhen-dong Zhang and Tariq Alkhalifah.
\newblock High-resolution reservoir characterization using deep learning-aided
  elastic full-waveform inversion: The north sea field data examplemlefwi.
\newblock {\em Geophysics}, 85(4):WA137--WA146, 2020.

\bibitem{dhara2022elastic}
Arnab Dhara and Mrinal Sen.
\newblock Elastic-adjointnet: A physics-guided deep autoencoder to overcome
  crosstalk effects in multiparameter full-waveform inversion.
\newblock In {\em SEG/AAPG International Meeting for Applied Geoscience \&
  Energy}. OnePetro, 2022.

\bibitem{wu2021cnn}
Yulang Wu, George~A McMechan, and Yanfei Wang.
\newblock Cnn-based gradient-free multiparameter reflection full-waveform
  inversion.
\newblock In {\em First International Meeting for Applied Geoscience \&
  Energy}, pages 1369--1373. Society of Exploration Geophysicists, 2021.

\bibitem{xu2022simultaneous}
Linan Xu, Zhaoqi Gao, Sichao Hu, Jinghuai Gao, and Zongben Xu.
\newblock Simultaneous inversion for reflectivity and q using nonstationary
  seismic data with deep-learning-based decoupling.
\newblock {\em IEEE Transactions on Geoscience and Remote Sensing}, 60:1--15,
  2022.

\bibitem{zhang2020numerical}
Tianze Zhang, Kristopher~A Innanen, Jian Sun, and Daniel~O Trad.
\newblock Numerical analysis of a deep learning formulation of multi-parameter
  elastic full waveform inversion.
\newblock In {\em SEG International Exposition and Annual Meeting}. OnePetro,
  2020.

\bibitem{yao2023regularization}
Jiashun Yao, Michael Warner, and Yanghua Wang.
\newblock Regularization of anisotropic full-waveform inversion with multiple
  parameters by adversarial neural networks.
\newblock {\em Geophysics}, 88(1):R95--R103, 2023.

\bibitem{deng2022openfwi}
Chengyuan Deng, Shihang Feng, Hanchen Wang, Xitong Zhang, Peng Jin, Yinan Feng,
  Qili Zeng, Yinpeng Chen, and Youzuo Lin.
\newblock Openfwi: Large-scale multi-structural benchmark datasets for full
  waveform inversion.
\newblock {\em Advances in Neural Information Processing Systems},
  35:6007--6020, 2022.

\bibitem{wu2019inversionnet}
Yue Wu and Youzuo Lin.
\newblock {InversionNet}: An efficient and accurate data-driven full waveform
  inversion.
\newblock {\em IEEE Transactions on Computational Imaging}, 6:419--433, 2019.

\bibitem{zhang2020data}
Zhongping Zhang and Youzuo Lin.
\newblock Data-driven seismic waveform inversion: A study on the robustness and
  generalization.
\newblock {\em IEEE Transactions on Geoscience and Remote sensing},
  58(10):6900--6913, 2020.

\bibitem{feng2023simplifying}
Yinan Feng, Yinpeng Chen, Peng Jin, Shihang Feng, Zicheng Liu, and Youzuo Lin.
\newblock Simplifying full waveform inversion via domain-independent
  self-supervised learning.
\newblock {\em arXiv preprint arXiv:2305.13314}, 2023.

\bibitem{levander1988fourth}
Alan~R Levander.
\newblock Fourth-order finite-difference p-sv seismograms.
\newblock {\em Geophysics}, 53(11):1425--1436, 1988.

\bibitem{christensen1996poisson}
Nikolas~I Christensen.
\newblock Poisson's ratio and crustal seismology.
\newblock {\em Journal of Geophysical Research: Solid Earth},
  101(B2):3139--3156, 1996.

\bibitem{gercek2007poisson}
H~Gercek.
\newblock Poisson's ratio values for rocks.
\newblock {\em International Journal of Rock Mechanics and Mining Sciences},
  44(1):1--13, 2007.

\bibitem{marfurt1984accuracy}
Kurt~J Marfurt.
\newblock Accuracy of finite-difference and finite-element modeling of the
  scalar and elastic wave equations.
\newblock {\em Geophysics}, 49(5):533--549, 1984.

\bibitem{engquist1977absorbing}
Bj{\"o}rn Engquist and Andrew Majda.
\newblock Absorbing boundary conditions for numerical simulation of waves.
\newblock {\em Proceedings of the National Academy of Sciences},
  74(5):1765--1766, 1977.

\bibitem{loshchilov2018decoupled}
Ilya Loshchilov and Frank Hutter.
\newblock Decoupled weight decay regularization.
\newblock In {\em International Conference on Learning Representations}, 2018.

\bibitem{brantut2019influence}
Nicolas Brantut and Emmanuel~C David.
\newblock Influence of fluids on vp/vs ratio: increase or decrease?
\newblock {\em Geophysical Journal International}, 216(3):2037--2043, 2019.

\bibitem{wang2012high}
X-Q Wang, Alexandre Schubnel, Jerome Fortin, EC~David, Yves Gu{\'e}guen, and
  H-K Ge.
\newblock High vp/vs ratio: Saturated cracks or anisotropy effects?
\newblock {\em Geophysical Research Letters}, 39(11), 2012.

\bibitem{alam2017near}
Md~Iftekhar Alam and Priyank Jaiswal.
\newblock Near surface characterization using vp/vs and poisson's ratio from
  seismic refractions.
\newblock {\em Journal of Environmental \& Engineering Geophysics},
  22(2):101--109, 2017.

\end{thebibliography}

\end{document}